\documentclass{jfm}
\usepackage{newtxtext}
\usepackage{newtxmath}
\usepackage{graphicx}
\graphicspath{{figures/}}

\usepackage{natbib}
\usepackage{hyperref}
\hypersetup{
	colorlinks = true,
	urlcolor   = blue,
	citecolor  = blue,
}

\usepackage{lineno}

\usepackage{xcolor}

\newcommand\DX[1]{\color{black} #1\color{black}}
\newcommand{\bds}{\boldsymbol}

\usepackage{booktabs}
\usepackage{subfigure}
\usepackage{multirow}
\usepackage{array}
\newcolumntype{P}[1]{>{\centering\arraybackslash}p{#1}}

\title[Reinforcement-learning-assisted control of four-roll mills]
{Reinforcement-learning-assisted control of four-roll mills: geometric symmetry and inertial effect}

\author[X. Dai, D. Xu, M. Zhang, Y. Yang]%
{
	Xuan Dai$^1$, Da Xu$^1$, Mengqi Zhang$^1$\thanks{Email address for correspondence: mpezmq@nus.edu.sg}, Yantao Yang$^2$
}
\affiliation{
	$^1$Department of Mechanical Engineering, National University of Singapore, 9 Engineering Drive 1, 117575 Singapore \\
	$^2$Department of Mechanics and Engineering Science, State Key Laboratory for Turbulence and Complicated System, College of Engineering, Peking University, 100871, People' Republic of China
	\\[\affilskip]
}

\date{\today}

\begin{document}
\maketitle

\begin{abstract}
Embedding the intrinsic symmetry of a flow system in training its machine learning algorithms has become a significant trend in the recent surge of their application in fluid mechanics. This paper leverages the geometric symmetry of a four-roll mill (FRM) to enhance its training efficiency. Stabilizing and precisely controlling droplet trajectories in a FRM is challenging due to the unstable nature of the extensional flow with a saddle point. Extending the work of Vona \& Lauga (\textit{Phys. Rev. E}, 104, 5, 055108, 2021), this study applies Deep Reinforcement Learning (DRL) to effectively guide a displaced droplet to the center of the FRM. Through direct numerical simulations, we explore the applicability of DRL in controlling FRM flow with moderate inertial effects, i.e., Reynolds number $\sim\mathcal{O}(1)$, a nonlinear regime previously unexplored. The FRM's geometric symmetry allows control policies trained in one of the eight sub-quadrants to be extended to the entire domain, reducing training costs. Our results indicate that the DRL-based control method can successfully guide a displaced droplet to the target center with robust performance across various starting positions, even from substantially far distances. \DX{The work also highlights potential directions for future research, particularly focusing on efficiently addressing the delay effects in flow response caused by inertia. } This study presents new advances in controlling droplet trajectories in more nonlinear and complex situations, with potential applications to other nonlinear flows. The geometric symmetry used in this cutting-edge reinforcement learning approach can also be applied to other control methods.
\end{abstract}

\begin{keywords}
Reinforcement learning, Flow control, Extensional flow, Four-roll mill
\end{keywords}

\section{Introduction}
In the seminal work by \cite{taylor1934formation}, a two-dimensional fluid system, now termed four-roll mill (FRM), was designed to analyse the deformation of drops and the formation of emulsions. In this setup, four identical cylinders submerged in a viscous liquid are driven by electric motors. By adjusting the rotational speeds of the rollers, the flow can vary from purely extensional to shear-dominated, to purely rotational. The attributes of FRM have made it popular in various applications. {\color{black} For example, the four-roll mill or similar device has been utilized to generate controlled extensional flows, facilitating the study of droplet deformation and suspension dynamics in microfluidic environments \citep{lee2007microfluidic,Hudson2004}, allowing for precise manipulation of cells, particles, and drops, which is essential for applications in material science and chemical engineering \citep{rumscheidt1961particle,bentley1986experimental}. The FRM has also been instrumental in studying the behaviour of polymer solutions under various flow conditions. By adjusting the flow type and rate, researchers can investigate polymer chain stretching and orientation, which are critical for optimising industrial processes like extrusion and moulding. Relevant works include \cite{fuller1981flow,feng1997numerical,Mackley2010}}. Notably, \cite{bentley1986computer} designed a computer-controlled FRM that is capable of producing arbitrary linear flow fields. An automated control mechanism was proposed to stabilise droplets in the centre of the flow cell. \cite{higdon1993kinematics} systematically investigated the extensional and rotational rates under different combinations of characteristic length ratios in a square box based on a two-dimensional simulation. For detailed reviews of the application of FRM in fluid mechanics, see \cite{Rallison1984,Stone1994}. More recently, \cite{vona2021stabilizing} utilised reinforcement learning (RL) to search for an optimal control policy that can drive a droplet to the center via modulation of roller speeds at vanishingly-small Reynolds number $Re$.

Given the unique importance of FRM in both academic research and real-world applications, it is of great interest to \DX{explore the accurate and robust control } of the droplet in the FRM. Previous papers by \cite{bentley1986computer,vona2021stabilizing} have laid a solid foundation. Based on these works, this study aims to extend the FRM control in the following two key aspects. First, we will consider moderate inertial effects with $Re\sim\mathcal{O}(1)$ in the FRM, a case seldom studied in the control of FRM. The effect of inertia on the control results will be elucidated in our task and this will help understand how the nonlinearity can be controlled in FRM. Second, we will leverage the geometric symmetry in FRM to facilitate the training and testing of the control policy. To the best of our knowledge, past works have not utilised the geometric symmetry in controlling the flow in FRM. 
Embedding and utilising intrinsic symmetry in machine learning algorithms represents a significant trend in the recent development \citep{vanderpol2021,Otto2023}.
With these improvements, our work aims to further test the applicability of deep RL (DRL) in guiding a droplet to the center of the FRM using a direct numerical simulation method. The reasons of choosing the DRL as the control method are twofold. First, \cite{bentley1986computer} demonstrated that a linear PID-type controller failed to stabilise a droplet, which will drift exponentially away from the stagnation point if uncontrolled. \DX{A PID-type controller regulates a process by combining proportional, integral, and derivative actions, reacting to errors and correcting past offsets in a linear manner. Its inability to control the extensional flow is likely due to the inherently linear nature of the controller, which may be insufficient for managing the complex, nonlinear dynamics of such a flow system. } Second, DRL has been applied successfully in controlling nonlinear flows \citep{rabault2019artificial}. It also represents the state of the art in the application of machine learning algorithms in controlling the unstable extensional flow in FRM, as first explored by \cite{vona2021stabilizing}.

In the following, we will first introduce the flow problem and explain the numerical methods \DX{in Section 2}. The results will then be discussed and compared to those in \cite{vona2021stabilizing} \DX{in Section 3. The section also explains the advantage of leveraging the geometric symmetry in the FRM and discusses the effect of inertia. In the end, we conclude the work with some discussions. Five appendices provide additional information on the delay in flow response due to inertia, effect of thermal noise, global policy regarding different initial conditions, hyperparameter fine-tuning and effects of different state definitions. } Our code will be shared online upon the acceptance of the work.


%

\section{Problem formulation and numerical methods}\label{Methodology}
\subsection{Direct numerical simulation and validation}\label{Equation}
\begin{figure}
	\centering
	\subfigure{\includegraphics[width=0.45\textwidth,trim= 0 13 0 10,clip]{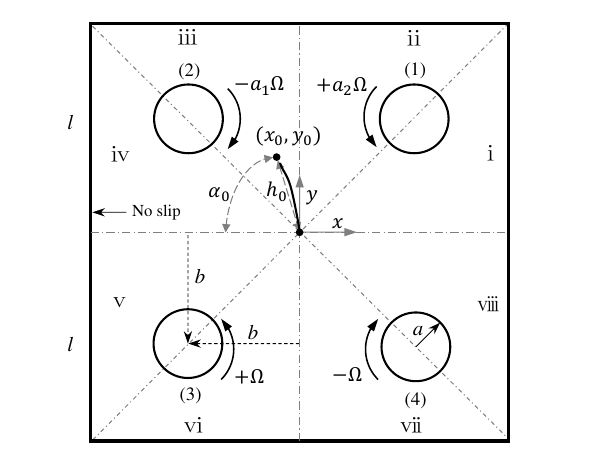}}
	\put(-170,120){$(a)$}
	\hspace{1.5cm}
	\put(-12,120){$(b)$}	
	\subfigure{\includegraphics[width=0.459\textwidth,trim= 0 0 0 0,clip]{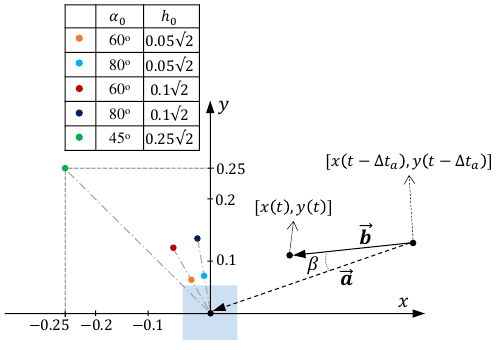}}
	\newline
	\subfigure{\includegraphics[width=0.63\textwidth,trim= 0 0 0 0,clip]{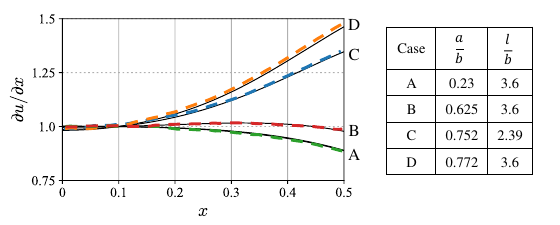}}
	\put(-260,90){$(c)$}	
	\caption{($a$) A schematic diagram of four-roll mill showing the control setup for a droplet initially positioned in the sub-quadrant ${\romannumeral 3}$. {\color{black} ($b$) The five initial positions of the droplet to be studied in the present work with different $h_0$ and $\alpha_0$. The blue shade encircles the initial positions considered by \cite{vona2021stabilizing}. The definition of the angle $\beta$ in the reward definition Eq. \ref{reward} is also illustrated. ($c$) Validation of our DNS results (black solid lines) against those (dashed lines) in \cite{higdon1993kinematics} at vanishingly small $Re$.} The letters represent the cases of different $a/b$ and $l/b$ as summarised in the table.}
	\label{FRM_instrument}
\end{figure}

The diagram in figure \ref{FRM_instrument}($a$) depicts the 2-D four-roll mill (FRM) instrument filled with a Newtonian fluid. The Cartesian coordinate originates from the center of a square domain with side length $2l$. The domain is divided into 8 sub-quadrants, which will be discussed in the section on the geometric symmetry. Positioned at $(\pm b, \pm b)$ are four rollers, each with a radius $a$. The rollers are indexed (1) through (4), corresponding to the first through the fourth quadrants, respectively. The baseline rotation rate of the rollers is denoted by $\pm\Omega$, with the positive (negative) sign representing the \DX{anticlockwise (clockwise) } direction. To generate an extensional flow, the baseline rotation rates of the rollers from (1) to (4) are designated as \DX{${\omega}_B=[+\Omega,\ -\Omega,\ +\Omega,\ -\Omega]$}, respectively. A droplet is initially positioned at $(x_0,y_0)$ in Cartesian coordinate or $(h_0,\alpha_0)$ in polar coordinate, where $\alpha_0$ denotes the initial angle between the droplet and the negative $x$-axis and $h_0$ is the initial radial distance. The two coordinates are related by $x_0=-h_0\cos(\alpha_0)$ and $y_0=h_0\sin(\alpha_0)$. Following \cite{vona2021stabilizing}, we make an assumption that the droplet is represented as a rigid fluid particle, meaning that it will not deform. Besides, the passive droplet experiences no external forces and its movement will not affect the flow field.

Although flow with $Re\sim\mathcal{O}(1)$ can be approximated as Stokes flow, to faithfully capture the inertial effect, we solve the 2-D dimensionless incompressible Navier-Stokes equations, contrary to the linear framework adopted previously \citep{bentley1986computer,vona2021stabilizing}, 
\begin{equation}
	\frac{\partial \boldsymbol{u}}{\partial t}+\boldsymbol{u} \cdot \nabla \boldsymbol{u}=-\nabla p+\frac{1}{R e} \nabla^{2} \boldsymbol{u}, \quad \nabla \cdot \boldsymbol{u}=0
	\label{eq3}
\end{equation}
where $\boldsymbol{u}=(u, v)^{T}$ denotes the velocity, $p$ the pressure and $Re=\frac{b\Omega a}{\nu}$, where $\nu$ is kinematic viscosity. Our length scale is $b$, velocity scale is $\Omega a$, \DX{time scale is $\frac{b}{\Omega a}$ } and pressure scale is $\rho \Omega^2 a^2$. When $Re$ is small, \DX{the convective terms can be neglected } and analytical solutions exist for the induced Stokes flow, \DX{as adopted in \cite{vona2021stabilizing}. In our study, however, we will retain all the terms even though the Reynolds number is small. This will facilitate the investigation of the (weak) inertial effect. Specifically, we will consider $Re=10^{-9}, 0.4, 2$ and 3. According to the data in \cite{bentley1986computer}, it is possible to realise these Reynolds numbers in experiments by adjusting the roller rotation rate and choosing a proper fluid.
For simplicity, the descriptions in this section will be based on the representative case $Re=0.4$, as they remain the same for the other cases unless otherwise noted.}


The roller rotation is realised by imposing a velocity boundary condition on the rollers. On the boundary of square domain, we consider no-slip boundary conditions following \cite{higdon1993kinematics}. To solve Eq. \ref{eq3}, the open-source code Nek5000 based on the spectral element method \citep{fischer2017nek5000} is utilised.
We have verified the mesh convergence of our FRM simulations, and chose a mesh composed of 1436 elements of the order 7 for a good balance between accuracy and computational cost. Regarding time integration, the two-step backward differentiation scheme is adopted with a time step $\Delta t=10^{-3}$ time units.


The past works \citep{fuller1980measurement,fuller1981flow,higdon1993kinematics} have identified $\frac{a}{b}$ and $\frac{l}{b}$ as two principle design parameters for determining proper approximation of extensional flow in the FRM test region. {\color{black} We have validated our numerical simulations at angular velocity amplitude of all rollers $\Omega=1.6\ rad/s$ within a range of $\frac{a}{b}$ and $\frac{l}{b}$ against the results in \cite{higdon1993kinematics} at vanishingly small $Re$, as shown in  figure \ref{FRM_instrument}($b$)}. More precisely, {\color{black} for the case $B$} of $\frac{a}{b}=0.625$ and $\frac{l}{b}=3.6$, \cite{higdon1993kinematics} reported that the extension rate at the origin under extensional flow is 0.7064 and the vorticity at the origin under rotational flow is 0.8250. Our numerical results yield 0.7065 and 0.8250 respectively. This case is chosen for the DRL control in our work. 



\subsection{Control setup}\label{Control_setup}

\begin{table}
	\begin{center}
		\scalebox{1}{
			\begin{tabular}{P{1cm}P{1.9cm}P{1.2cm}P{1.2cm}|P{0.5cm}P{0.5cm}P{0.5cm}P{0.5cm}P{1cm}P{0.5cm}P{1.3cm}P{0.5cm}}
		Case & Re &	$h_0$ & $\alpha_0$ & $\eta$ & $p$ & $q$ & $c$ & $h_{e}$ & $\gamma$ & $\Delta t_a$ &$N$ \\
		1,2	& 0.4 &	0.05$\sqrt{2}$ & $60^\circ, 80^\circ$ & 1.5 & 2 & 30 & 2 & 0.0025 & 0.98 & 0.05 & 30 \\
		3,4	& 0.4 &	0.1$\sqrt{2}$ & $60^\circ, 80^\circ$ & 2 & 2 & 10 & 8 & 0.005 & 0.99 & 0.05 &50 \\
		\multicolumn{1}{c}{\color{black}5.1, 5.2, 5.3}	& [$10^{-9}$,0.4,2] & 0.25$\sqrt{2}$ & $45^\circ$ & 3 & 2 & 7 & 10 & 0.005 & 0.99 & 0.05 &90\\
		{\color{black}5.4} & 3 & 0.25$\sqrt{2}$ & $45^\circ$ & 3 & 2 & 7 & 10 & 0.005 & 0.99 & 0.075 &90
			\end{tabular}
		}
		\caption{{\DX{The cases considered in this work and the parameters selected in each case under $a/b=0.625$, $l/b=3.6$. $h_0$: initial radial distance to origin; $\alpha_0$: initial angle; $\eta$: clipped value for sampling actions; $p,q,c$: parameters in the reward function Eq. (\ref{reward}); $h_e$: target distance to origin; $\gamma$: discounting factor; $\Delta t_a$: time interval between adjacent control actions; $N$: {\color{black}maximum control steps per epoch}.}}} 
		\label{Train_parameters}
	\end{center}
\end{table}

\begin{table}
	\begin{center}
		\scalebox{1}{
			\begin{tabular}{P{3cm}P{3cm}P{3cm}P{3cm}}
				$h_0$ & $\alpha_0$ & $\eta$ & $p,q,c$  \\
			    Initial distance & {\color{black} Initial angle} & Clipped value for sampling actions & Parameters in reward function  \\
			    \\
			    $h_e$ & $\gamma$ & $\Delta t_a$ & $N$  \\
			    Target distance & Discounting factor & Time interval between actions & {\color{black} Maximum control steps per epoch}  \\
			\end{tabular}
		}
		\caption{{\DX{Explanations for the parameters in Table\ \ref{Train_parameters}.}}} 
		\label{Meaning_parameters}
	\end{center}
\end{table}

In \cite{vona2021stabilizing}, a rotlet solution of the flow induced by the rotation of rollers was assumed in an idealised Stokes flow, where the drop could respond to the changes of roller speed instantaneously. In their control setup, they controlled the rotation rate of one roller with the other three rotating at a default angular velocity. However, in our simulations, we discovered that adjusting only one roller is overly restrictive and inefficient in the finite-$Re$ regime due to the existence of non-negligible inertia. As a result, our focus shifted to actuating two adjacent rollers closest to the initial position of the droplet. \DX{Note that for the case of $Re=3$, control using three rollers is necessary. For clarity, the following explanation of the control setup will focus on two rollers, with the expansion to three rollers discussed later in \mbox{Sec.\ \ref{result_Re3}}.}

The chosen rollers will not change within a single control task.
For example, given ${\omega}_B$, for a droplet initially placed in the sub-quadrant $iii$, the two adjacent rollers chosen to control its trajectory are roller (1) and roller (2), see figure \ref{FRM_instrument}($a$). We will modulate the baseline rotation rate of the roller closest to the droplet by multiplying it with an adjustable signal $a_1(t)$ and similarly for the one less close, multiply it by $a_2(t)$. The DRL-controlled rotation rates of the rollers in this case then become ${\omega}^{\romannumeral 3}(t)=[+a_2(t)\Omega,\ -a_1(t)\Omega,\ +\Omega,\ -\Omega]$. The DRL algorithm aims to determine the signals $a_1(t),a_2(t)$ in time, to be elucidated shortly.

{\color{black} Five different initial positions of the droplet are considered, see figure \ref{FRM_instrument}($b$).} Among them, four initial positions are located within the sub-quadrant $iii$ with varying distances to the origin and angles, i.e., $[h_0,\alpha_0]=[0.1\sqrt{2},\ 60^{\circ} \text{ or } 80^{\circ}]$, $[0.05\sqrt{2},\ 60^{\circ} \text{ or } 80^{\circ}]$. In \cite{vona2021stabilizing}, the initial positions of the drops were confined in the region of $x\in(-0.05,0.05),y\in(-0.05,0.05)$ (they normalised the length in the same way as we did). Thus, their drops were placed within $0.05\sqrt{2}$ of the origin. It is also noted that their normalised radius of the rollers is 0.8, which is slightly greater than ours. 
Our fifth case with $[h_0,\alpha_0]=[0.25\sqrt{2},\ 45^{\circ}]$ defines a challenging task since the droplet is positioned substantially far away from the origin. \DX{For this case, we also vary the $Re$ to investigate the inertial effect.}

\subsection{Deep reinforcement learning}\label{drl}
DRL is a machine-learning algorithm that leverages deep learning techniques and reinforcement learning principles to automate the decision-making process \citep{Brunton2020}.
 The core concept of DRL-based control relies on an agent, approximated by an artificial neural network, learning to identify the optimal control policy through continuous interaction with the environment. By assessing the outcomes of its actions as either desirable or undesirable, the agent learns and adapts from these experiences according to a user-defined reward function.



The state input in our DRL algorithm includes the droplet's position, velocity and acceleration, together defined as $s_t=[x(t),\ y(t),\ u(t),\ v(t),\ k_x(t),\ k_y(t)]\in\mathcal{S}$. The position is obtained by integrating the velocity signal and the acceleration is calculated by differentiating the velocity signal in time. In contrast, \cite{vona2021stabilizing} used solely the position as the state. \DX{In our simulations, using }only the position as the input failed in the finite-$Re$ regime, which may be related to the non-negligible inertial effect in our case. The actions at time $t$ are $a_t=[a_1(t),\ a_2(t)]\in\mathcal{A}$, adjusting the baseline rotation rates of the two adjacent rollers as explained earlier. The values of $a_1(t),\ a_2(t)$ are \DX{sampled } between $[-\eta,\ \eta]$, where $\eta$ is a predefined constant. The reward function $r(t)\in\mathbb{R}^+$ consists of $r_1(t)$, $r_2(t)$ and $r'$, i.e.,
\begin{equation}\label{reward}
	\begin{aligned}
		r(t) & = r_1(t) + r_2(t)+r'  =\exp [-p(1-\cos \beta(t))]+\exp [-q h(t)]+r'. \\
	\end{aligned}
\end{equation}
The definition of the reward $r_1(t)$ follows the work of \cite{vona2021stabilizing} and is related to $\beta(t)$, defined as the angle between the displacement vector $\vec{b}=\left[x(t)-x(t-\DX{\Delta t_a}),\ y(t)-y(t-\DX{\Delta t_a})\right]$ and the inward vector $\vec{a}=\left[-x(t-\DX{\Delta t_a}),\ -y(t-\DX{\Delta t_a})\right]$, {\color{black} as illustrated in figure \ref{FRM_instrument}($b$),} where $\cos\beta(t)=\frac{x(t-\DX{\Delta t_a})\left[ x(t-\DX{\Delta t_a})-x(t)\right]+y(t-\DX{\Delta t_a})\left[ y(t-\DX{\Delta t_a})-y(t)\right] }{h(t-\DX{\Delta t_a}) \sqrt{\left[x(t)-x(t-\DX{\Delta t_a})\right]^2+\left[y(t)-y(t-\DX{\Delta t_a})\right]^2}}$ and $h(t)=\sqrt{x(t)^2+y(t)^2}$ and \DX{$\Delta t_a$ is the time interval for updating actions, i.e. the time interval between two control steps. }
The function $r_2(t)$ measures the droplet's radial distance to the origin. The last term $r'$ is defined as $
	r'=\left\{
	\begin{aligned}
		c & , &h(t) \leq h_e \\
		0 & , &h(t)>h_e
	\end{aligned}
	\right.$ 
which is relevant only when the droplet reaches the target radial distance denoted by $h_e$ and $c$ is the final reward for the droplet reaching the target. 
\cite{vona2021stabilizing} solely utilised $r_1(t)$ in their reward function, which did not work well in our experiments of finite-$Re$ flows. Thus, we added $r_2(t)$ to further incentivise the DRL controller to continuously increase the rewards as $h(t)$ decreases and also $r'$ for the terminal reward. 
In the above definition, $p,q$ are user-defined constants.
A parametric study on $p$ has been conducted by \cite{vona2021stabilizing}, which indicates that $p=[0.5,\ 1,\ 1.5,\ 2]$ do not differ significantly and they chose $p=1$. In our DRL training, we found that $p$ can affect the convergence of training and the stability of the policy. \DX{As detailed in Appendix \ref{hyper}, we studied the effect of hyperparameters in the reward functions and } set $p=2$ in order to encourage the droplet to move in the direction pointing to the origin. {\color{black} The values of the aforementioned parameters are summarised in table \ref{Train_parameters} with their meanings explained in table \ref{Meaning_parameters}.}

\DX{The time interval between two consecutive actions, denoted as $\Delta t_a$, appears to be an important parameter. Its effect on the flow control can be similarly studied as in \cite{bentley1986computer} by varying the time interval. We will consider fixed values of $\Delta t_a$. Specifically, the action is updated every 50 time steps, or $\Delta t_a=0.05$, in the cases $Re=[10^{-9},0.4,2]$, whereas the action is updated every 75 time steps, or $\Delta t_a=0.075$, in the case of $Re=3$, as shown in Table \ref{Train_parameters}). \mbox{Appendix \ref{appendix_ta}} shows a heuristic approach for determining $\Delta t_a$. } {\color{black} Unlike in \cite{vona2021stabilizing}, wherein actions ramp up to their new values on a finite timescale, the actions here are constantly applied and unchanged during a control step, until it is updated at the next control step.}

Proximal policy optimisation (PPO) is utilised as the training algorithm, which has a typical policy-based actor-critic network structure \citep{Sutton2018}. The actor network $\pi_\theta\left(a_t \mid s_t\right)$ takes the state $s_t$ as the input and generates a probability distribution from which the action $a_t$ is sampled. The critic network $V_\phi(s_t)$ predicts the value function of state $s_t$, i.e. the discounted rewards starting from the state $s_t$. The critic aims to provide an accurate prediction to minimise the objective function. For more technical details on PPO, the reader is referred to \cite{schulman2017proximal}. This method has been used in previous DRL works applied to the flow control problems \citep{rabault2019artificial} and also in our past work \citep{li2022reinforcement}. Both the actor and the critic networks consist of 2 hidden layers each with 300 neurons using ReLU as the activation function. The networks are updated using the Adam-optimiser \citep{kingma2014adam} with the learning rate of 0.0001 and of 0.0002 respectively. The allowable steps $N$ in each epoch and the discount factor $\gamma$ are case-dependent, as summarised in table \ref{Train_parameters}. 

{\color{black} It is worthy mentioning that in addition to PPO, there are other algorithms in RL control. Based on how the DRL algorithms collect and utilize data, they can generally be classified into two categories, i.e., on-policy and off-policy methods. On-policy algorithms, such as PPO, learn exclusively from data generated by the current policy being trained. While this ensures that the training data is always aligned with the policy, it can lead to lower sample efficiency since past experiences are discarded. Off-policy algorithms, such as SAC (Soft Actor-Critic) and DDPG (Deep Deterministic Policy Gradient), can learn from historical data collected by any policy, including those different from the current one. This reuse of past experiences makes off-policy methods more sample-efficient. However, their off-policy nature often introduces challenges in stability and convergence, requiring careful hyperparameter tuning and additional optimization techniques.

In the fluid dynamics community, PPO has been widely adopted due to its stability and robustness, as demonstrated in studies such as \cite{rabault2019artificial, fan2020reinforcement, ren2021applying, li2022reinforcement}. DDPG has also been successfully applied in works like \cite{bucci2019control, Zeng2021, Kim2022, xu2023reinforcement}. For this study, we opted for PPO primarily because of its stability and lower sensitivity to hyperparameter variations. Furthermore, our approach leverages geometric symmetry, which already enhances the sample efficiency. This reduces the importance of the trade-off between stability and sample efficiency, making PPO a suitable choice for our framework. While other DRL algorithms might offer performance improvements, we believe that the core novelty of our work, that is investigatinng inertial effects and geometric symmetry, is effectively captured within our current numerical framework. This will be demonstrated in the result section.
}

\DX{In the end, we explain the other theme of the work, that is how to utilise the geometric symmetry in FRM for DRL control. In general, leveraging symmetry to enhance models in machine learning has recently emerged as a prominent research trend \citep{Otto2023}. In DRL-related works, similar concepts have demonstrated improved model performance, such as the invariance of locality discussed in \cite{Belus2019, Vignon2023, Vasanth2024, Suarez2024, Suarez2024b}. In these works, the reward function is densely defined within specific local regions influenced by control actions. However, we emphasize that the geometric symmetry employed in this study is distinct from these approaches. It is derived from the global transformation framework introduced in \cite{vanderpol2021}. }
Thanks to the geometric symmetry of FRM, the whole domain can be evenly divided into 8 sub-quadrants from $i$ to $viii$, as shown in figure \ref{FRM_instrument}($a$). Following the notations in \cite{vanderpol2021}, for state $s\in\mathcal{S}$, action $a\in\mathcal{A}$ and reward $r\in\mathbb{R}^+$, under transformation operators $L_g:\mathcal{S}\rightarrow\mathcal{S}$ and $K_g^s:\mathcal{A}\rightarrow\mathcal{A}$, a symmetry-enhanced DRL algorithm can be constructed as 
\begin{equation}
{r}(s,a)={r}(L_g[s], K_g^s[a]),
\end{equation}
where $L_g$ or $K_g^s$ defines the transformation of state or action utilising the inherent symmetry. The equation implies that the immediate reward of the state-action pair remains the same after the transformation. So, $s$ (or $a$) and $L_g[s]$ (or $K_g^s[a]$) are equivalent in $\mathcal{S}$ (or $\mathcal{A}$). One can train a control policy in the lifted space (i.e., one of the 8 sub-quadrants) and, once the training process is converged, map the policy back and apply it to the entire domain \DX{(see section \ref{results_flow_symmetry} for details)}. Note that this concept is different from constraining the numerical simulations of FRM in a sub-quadrant.

\section{Results and discussion} \label{Results}

\begin{figure}
	\centering
	\subfigure{\includegraphics[width=1\textwidth,trim= 2 5 2 4,clip]{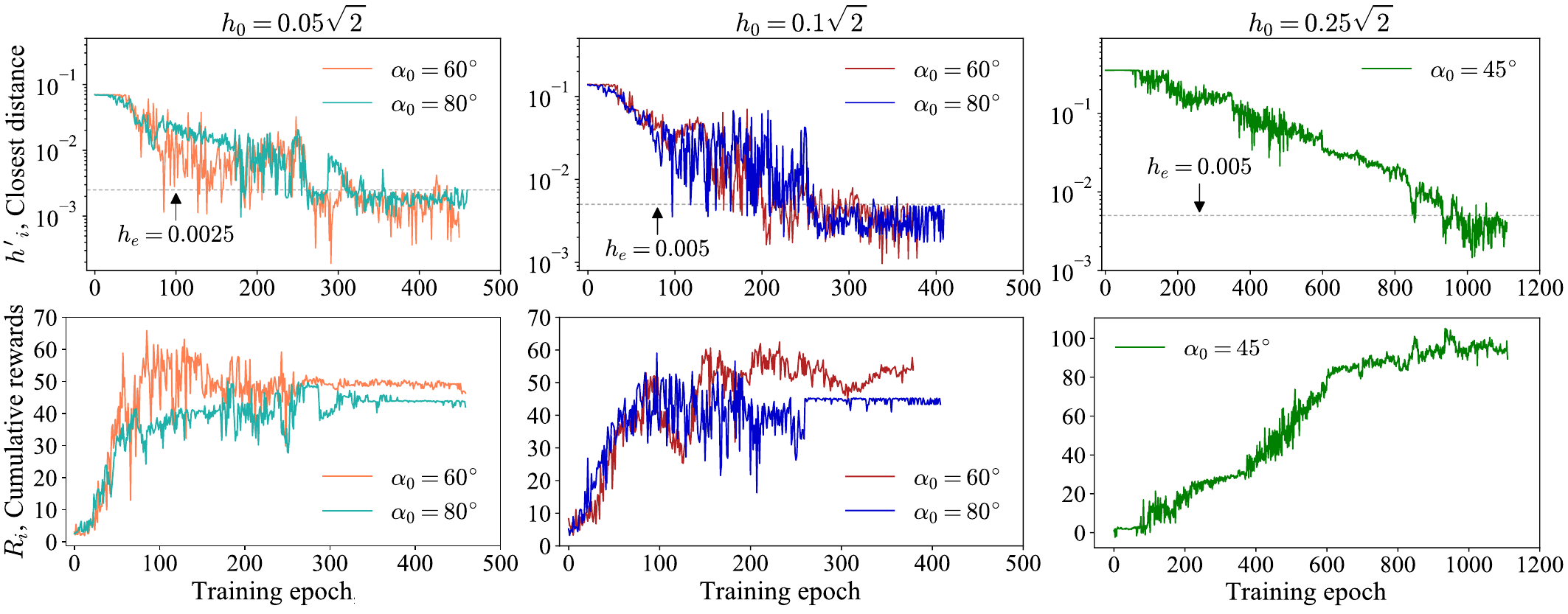}}
	\caption{Agent training history for droplets at different initial positions. The first row corresponds to the closest radial distance to the origin. The second row shows the cumulative rewards of each epoch.}
	\label{Train_hist}
\end{figure}

\DX{In the following, we will first present the typical training results of DRL for $Re=0.4$. Then, the DRL policy leveraging the geometric symmetry of FRM will be constructed to demonstrate the advantage of symmetry consideration. Finally, we will focus on the other theme of the paper, i.e., the characterisation of the inertial effect in the framework of DRL control. }

\subsection{Training results of DRL for $Re=0.4$}\label{Results_train_test}

This section explains the controlled results using the DRL method for a droplet initially placed in the sub-quadrant ${\romannumeral 3}$. Note that in this case it is the roller (1) and (2) which we implement the modulation of the rotation rates of. Figure \ref{Train_hist} displays the agent training history in terms of epochs for all the five considered cases. In the first row, $h'_i$ measures the minimum distance of the droplet to the origin in the $i$-th epoch. As the training proceeds, the value of $h'_i$ gradually decreases and consistently stays below $h_e$ in the end. During the training, there are occasional "downward overshoots" of $h'_i$ falling below $h_e$. We consider the training to have converged when there are more than 30 to 50 consecutive epochs of successful control, indicated by $h'_i<h_e$. Overall, $\sim$400 epochs are commonly required to train the cases 1-4 and $\sim$1100 epochs for the case 5. This difference additionally testifies the difficulty in controlling the last case, which is substantially far from the origin. The second row of the figure shows the cumulative rewards $R_i$, which adds up all the immediate rewards of the training steps in the $i$-th epoch. In the cases 1-4, the values of $R_i$ initially rise rapidly, followed by significant fluctuations, and eventually stabilise as the droplet nears the origin. The case 5 presents a continuous climbing-up trend in the training process. Combined together, the results of $h_i$ and $R_i$ imply that the agent learns from the interactions with the flow environment and updates its policy to guide the droplet to move towards the origin. With a sufficient amount of training epochs, the droplet is capable of reaching the target distance indicated by $h_e$. 
These results extend those in \cite{vona2021stabilizing}, where initial positions of drops typically within $h_0<0.05\sqrt{2}$ were considered, and demonstrate that droplets placed further away from the origin can also be controlled successfully by DRL.


\begin{figure}
	\centering
	\subfigure[Droplets trajectories in cases 1-4.]{\includegraphics[height=0.33\textwidth,trim= -20 10 0 7,clip]{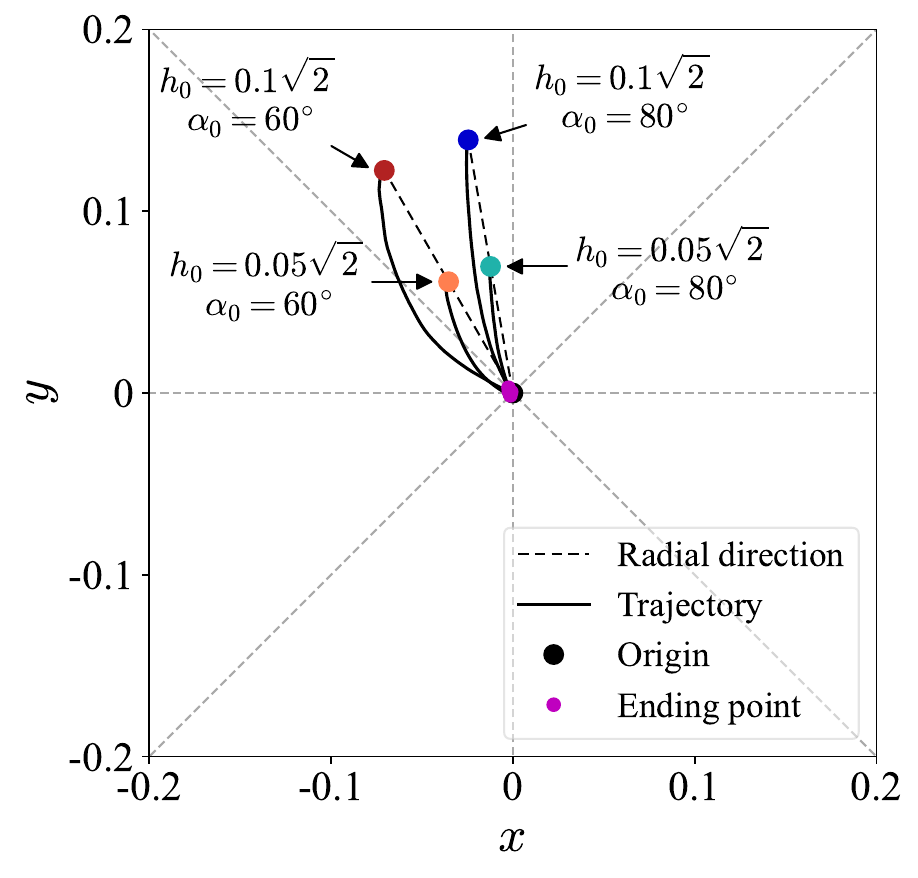}}
	\quad	\vspace{-0.35cm}
	\subfigure[{\DX{Modulations of roller actions for cases 1--4, where only two rollers are active: \(a_1(t)\) corresponds to roller (2), and \(a_2(t)\) corresponds to roller (1).}}]{\includegraphics[height=0.33\textwidth,trim= 0 6 0 7,clip]{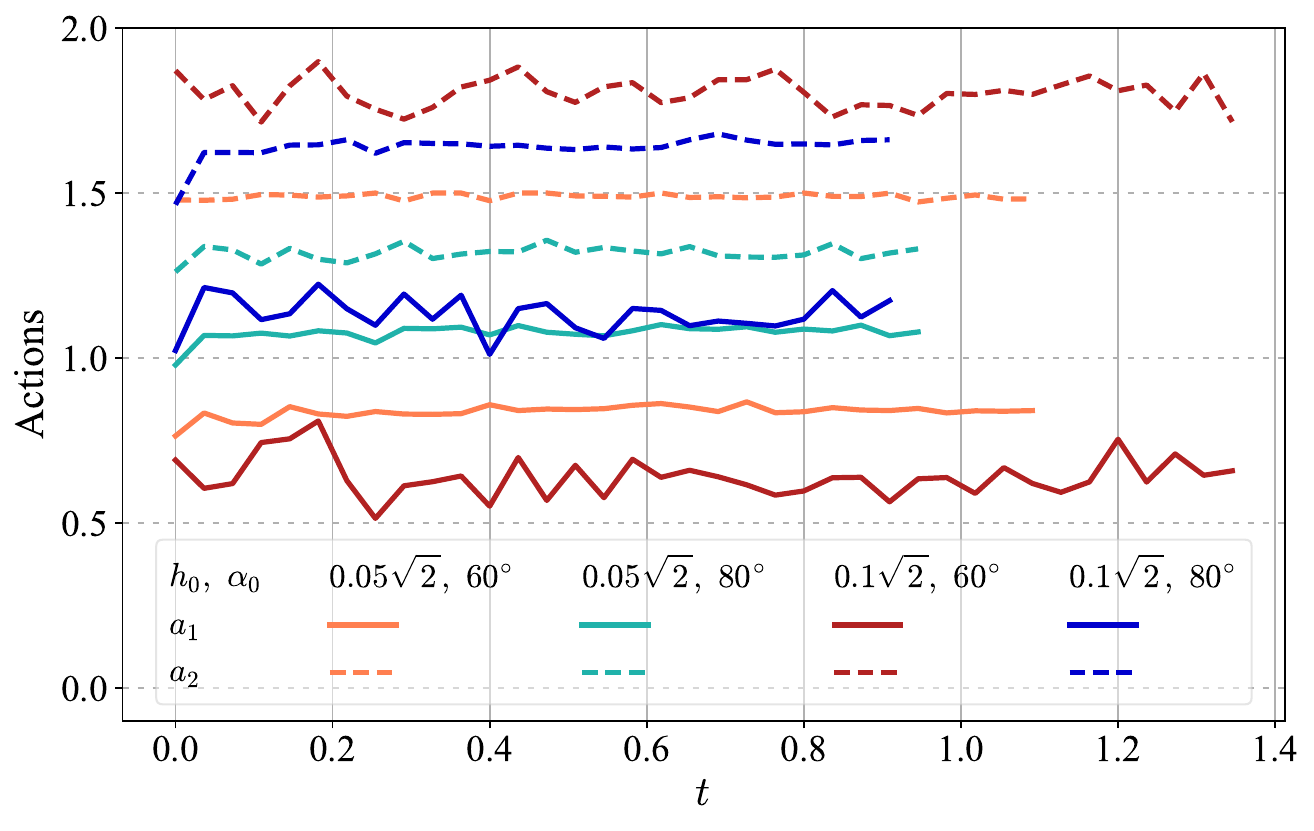}}
		\vspace{-0.2cm}
	\subfigure[Droplet trajectory in case {\color{black}5.2}.]{\includegraphics[height=0.33\textwidth,trim= -30 10 0 6,clip]{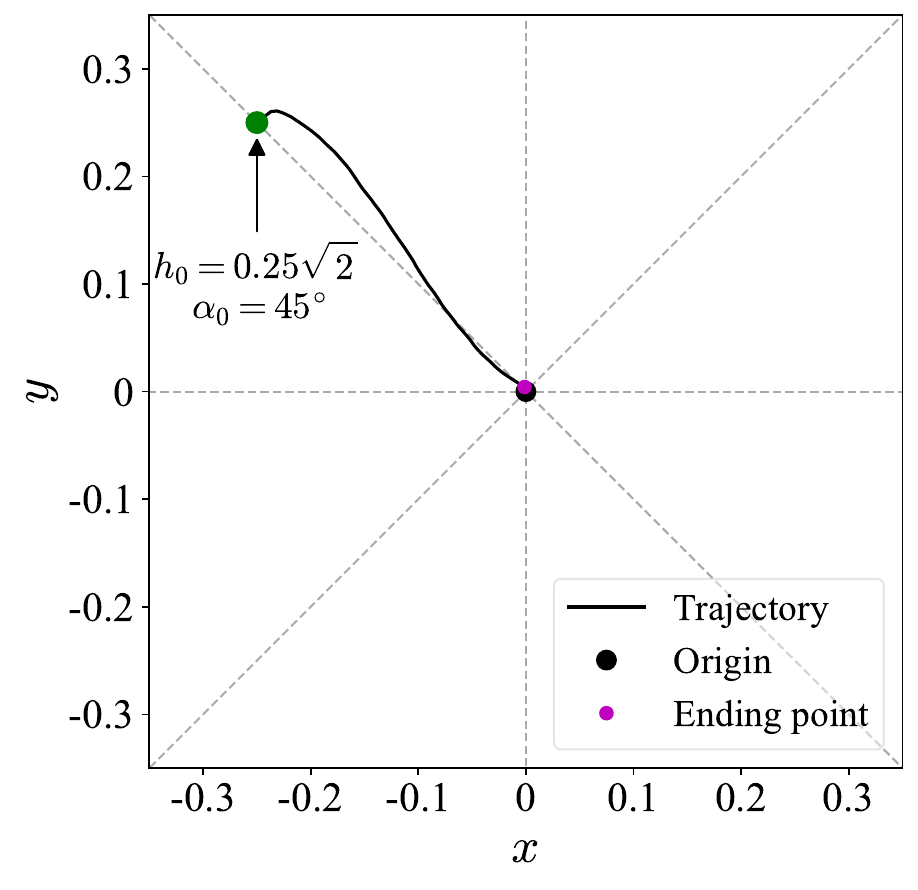}}
	\quad
	\subfigure[{\DX{Modulations of roller actions for case {\color{black}5.2}, where only two rollers are active: \(a_1(t)\) corresponds to roller (2), and \(a_2(t)\) corresponds to roller (1).}}]{\includegraphics[width=0.55\textwidth]{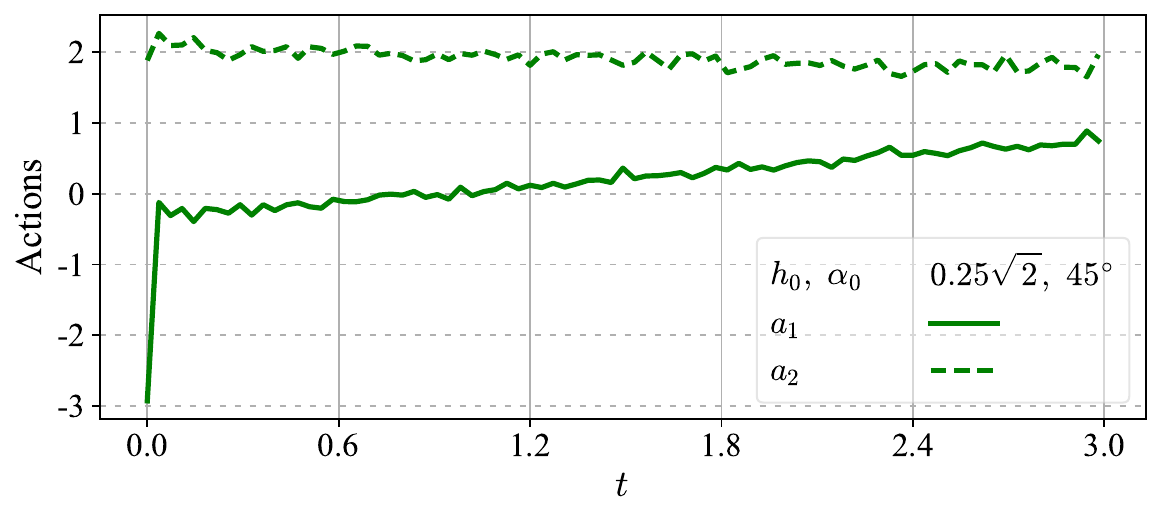}}
	\caption{Converged policies in figure \ref{Train_hist} are run for $50$ {\DX{episodes}} at $Re=0.4$, All droplets are successfully driven to the target radial distance. Panels (a), (c) demonstrate representative trajectories for all the considered cases. Panels (b), (d) display the corresponding actions. {\DX{Note that only two rollers (1) and (2) are acted on, and an action is followed by a waiting time $\Delta t_a=50\Delta t$ until the next one, during which it is unchanged. Therefore, in each control step shown in panel (b), (d), actions are actually step functions, and continuous lines in these panels are guides for the eye.}}}
	\label{Traj_r}
\end{figure}

To test the effectiveness of trained policies, the converged ones have been run for additional $50$ epochs and we find that in all the tests, the droplets can be driven back to the origin within the target radial distance $h_e$. Figure \ref{Traj_r} draws the representative trajectories and the associated actions for cases 1-4 in panels ($a,b$) and for case 5 in panels ($c,d$). It can be noticed that these trajectories exhibit smooth transitions from their starting point to the ending point, which are in contrast to the results in \cite{vona2021stabilizing} where the paths manifest zigzags. Possible reasons may include that \cite{vona2021stabilizing} studied a model of inertia-less Stokes flow without a time derivative term, while our work is based on DNS with finite inertial effect. Another possible reason might be that \cite{vona2021stabilizing} \DX{reset the roller velocity to its default right before the next action, giving rise to zig-zags}, whereas we continuously apply the action in all steps.

\begin{figure}
	\centering
	\subfigure{\includegraphics[height=0.48\textwidth,trim= 0 0 0 0,clip]{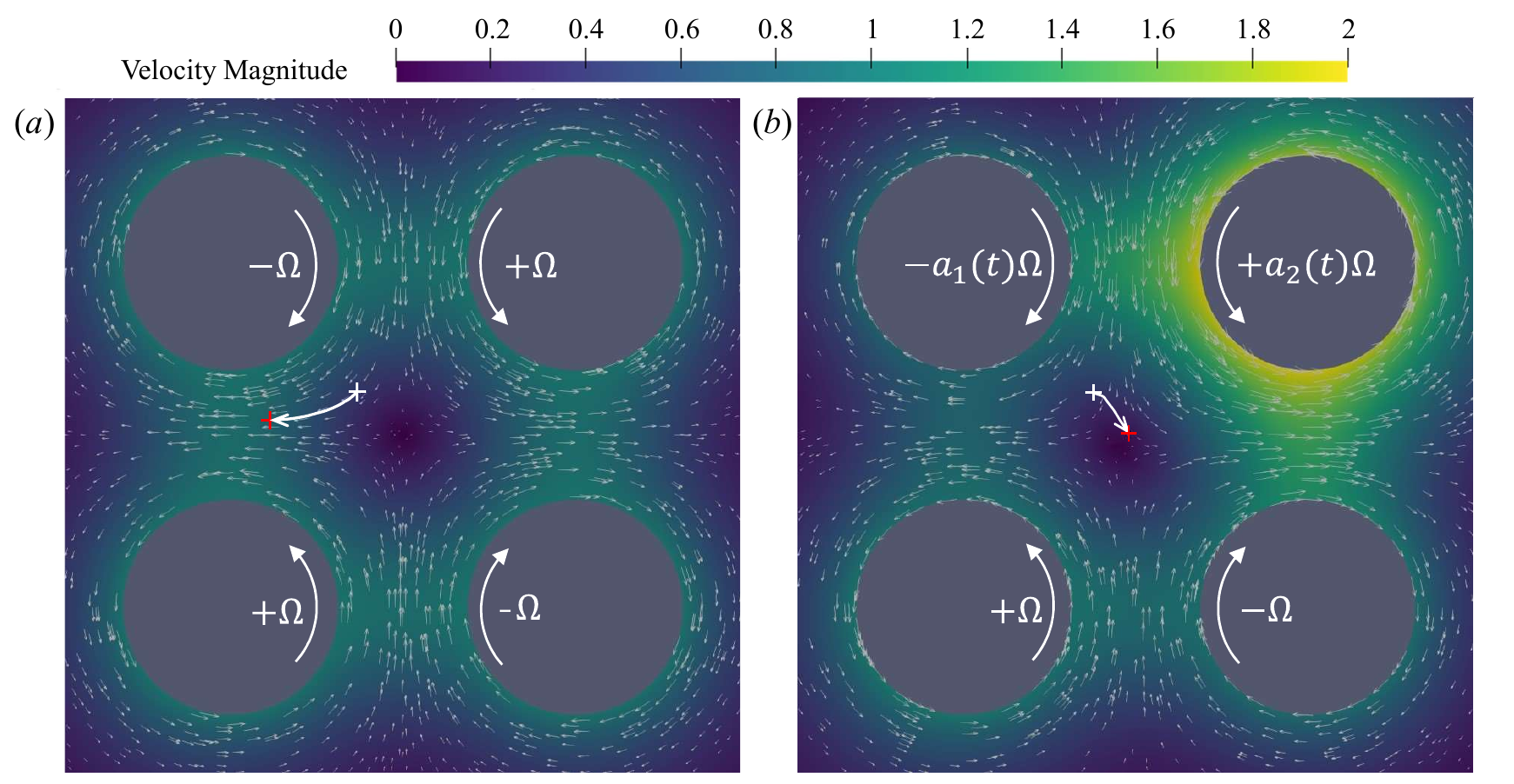}}
	\caption{Instantaneous quiver plots for the velocity field in case {\color{black}5.2} at $Re=0.4$. (a) Without control. The droplet {\DX{tends to be swept away}}. The flow is inherently symmetric. (b) The last time step of an epoch for a converged policy. The white cross represents the starting position of the droplet and the red one the ending position in this control case of  $h_0=0.25\sqrt{2}$ and $\alpha_0=45^{\circ}$.}
	\label{quiverplot}
\end{figure}

The actions exerted by the agent reflect a controlling logic which is consistent with the underlying flow physics. Specifically, the difference between $a_1$ and $a_2$ becomes larger with smaller $\alpha_0$, see figure \ref{Traj_r}($b$). Note that for all the considered droplets initiated in sub-quadrant $iii$, $a_1$ is assigned to roller (2) and $a_2$ is assigned to roller (1). The extensional flow generated by the baseline rotation rate ${\omega}_B$ presents influx from the top/bottom and outflux towards the left/right in the regions between the rollers \DX{(see figure \ref{quiverplot})}. The droplet initially positioned with smaller $\alpha_0$ tends to be \DX{swept } away by the outflux with a stronger left-pulling force; in order to counteract the left-pulling force exerted by the outflux, the roller (1) modulated by $a_2(t)$ should work "harder" to steer the droplet to move clockwise. Thus, the value of $a_2(t)$ is in general larger in the case of smaller $\alpha_0$ for the same $h_0$. This also explains why the DRL algorithm yields a controlled trajectory which deviates more from the straight direction (the dashed lines \DX{see figure \ref{Traj_r}($a$)}) in the smaller $\alpha_0$ cases. This effect is less severe in the cases of larger $\alpha_0$, resulting in a smaller difference between $a_1$ and $a_2$ in panel ($b$). In case {\color{black}5.2}, the sign of $a_1$ is even negative at the beginning (see figure \ref{Traj_r}$d$), reversing its rotation direction, which also aims to counteract the local outflux pulling the droplet to the left and, together with $a_2$, generate a trajectory as shown in panel \ref{Traj_r}($c$). Figure \ref{quiverplot} shows exemplary instantaneous quiver plots of the velocity field in {\color{black}case 5.2}. Without control, the droplet will move exponentially away from the initial position, as shown in panel ($a$). That the roller (1)'s action is stronger is also consistent with the roller choice in \cite{vona2021stabilizing}. The inertial effect will be further compared to small-$Re$ flows in Sec. \ref{vanishingRe}.

\DX{In the end, we would like to discuss the robustness of the trained policy under random noise and its generalisability to other initial conditions. Appendix \ref{result_noise} demonstrates the effectiveness of the policies in noisy environments by introducing a thermal noise term into the NS equation. Additionally, Appendix \ref{result_nearbyICs} explores the potential for a global policy by applying the policy trained for the specific initial condition \(\bds{x}_0 = (-0.03, 0.02)\) to other points in a nearby region. The results reveal that while droplets released from initial positions close to the one used for training the policy can sometimes be successfully controlled, the policy remains sensitive to initial conditions otherwise. Several factors may contribute to this sensitivity. First, in regions of extensional flow with steep flow gradients, small perturbations in the initial conditions can lead to significant trajectory deviations, complicating the control task. Second, since the DRL training is tailored to a specific initial position, the trained policy may not generalize well to regions with distinct flow characteristics, increasing the likelihood of control failure for significantly different initial conditions. Nonetheless, our findings indicate that the policy can effectively manage certain nearby initial positions, as shown by the green dots in figure \ref{Nearby_test} in Appendix \ref{result_nearbyICs}.
}



\subsection{DRL policy leveraging geometric symmetry of FRM}\label{results_flow_symmetry}

\begin{figure}
	\centering
	\subfigure{\includegraphics[width=0.45\textwidth,trim= 0 13 0 9,clip]{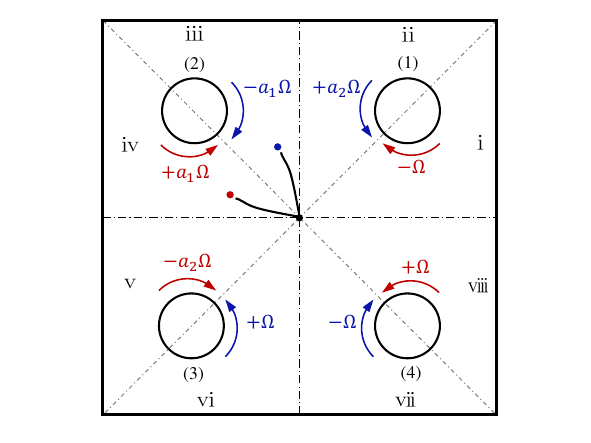}}
	\put(-159,106){$(a)$}
	\subfigure{\includegraphics[width=0.45\textwidth,trim= 0 13 0 9,clip]{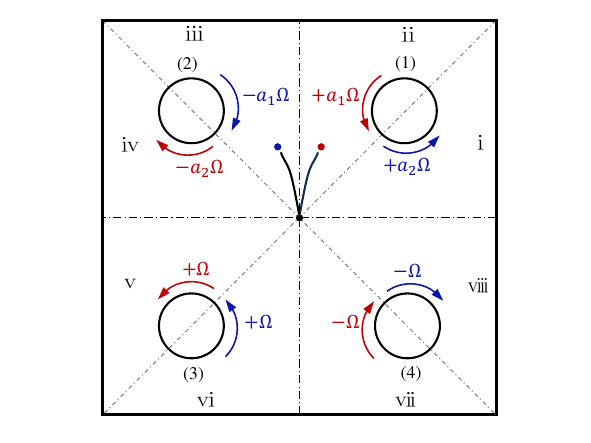}}
	\put(-159,106){$(b)$}	
	\caption{The domain of FRM is evenly divided into 8 sub-quadrants. By the geometric symmetry, the control policy trained in one of the sub-quadrants can be applied to the entire domain. For example, panel ($a$) shows that the dynamics of the blue droplet in sub-quadrant $iii$ and that of the red droplet in the sub-quadrant iv are symmetric with respect to the antidiagonal line (-$45^o$). Panel ($b$) shows that symmetry with respect to the vertical axis for the droplets in the sub-quadrants $ii$ and $iii$.}
	\label{FRM_act}
\end{figure}

In section \ref{Control_setup}, we have briefly mentioned that the geometric symmetry of FRM enables the control policies trained in one sub-quadrant to be applied to the entire flow domain. This section elaborates on the utilisation of this symmetry using policies trained in the last section \ref{Results_train_test}. To begin, considering a droplet initially positioned in the sub-quadrant ${\romannumeral3}$, represented by the blue dot in figure \ref{FRM_act}, we denote the state of the droplet at time $t$ as $s^{\romannumeral 3}(t)=[x(t),\ y(t),\ u(t),\ v(t),\ k_x(t),\ k_y(t)]$, and the actions determined by the agent at time $t$ as $a(t)=[a_1(t),\ a_2(t)]$ as in \DX{ ${\omega}^{\romannumeral3}(t)=[+a_2(t)\Omega,\ -a_1(t)\Omega,\ +\Omega,\ -\Omega]$}. 
The optimal policy is thus denoted as $\pi_\theta\Big(a(t) | s^{\romannumeral3}(t)\Big)$. By utilising the proper transformations based on the geometric symmetry, this policy $\pi_\theta$ can be applied to the entire domain, as explained below.

\begin{figure}
	\centering
	\subfigure{\includegraphics[height=0.28\textwidth,trim= 0 11 0 7,clip]{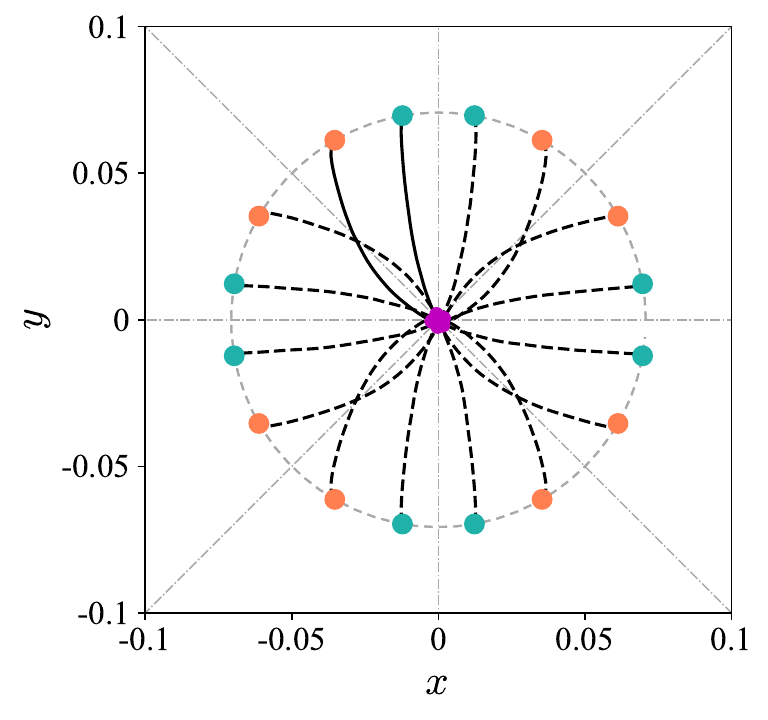}}
	\put(-120,100){$(a)$}	
	\quad
	\subfigure{\includegraphics[height=0.28\textwidth,trim= 0 11 0 7,clip]{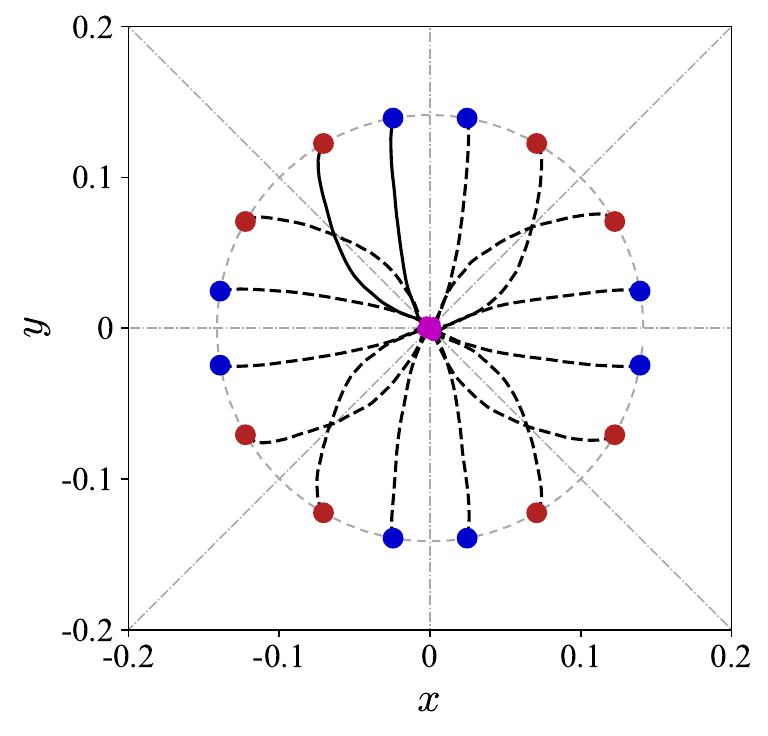}}
	\put(-120,100){$(b)$}	
        \quad	
	\subfigure{\includegraphics[height=0.28\textwidth,trim= 0 11 0 7,clip]{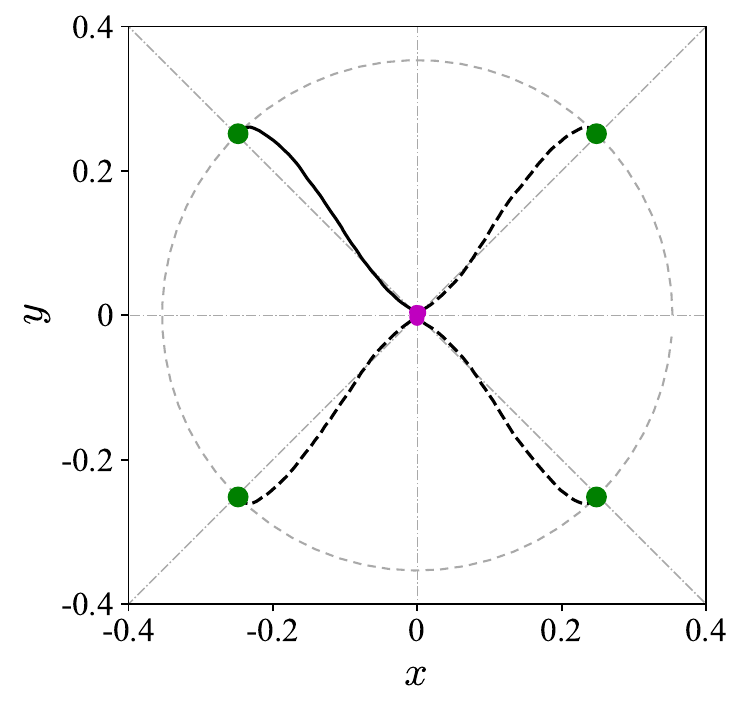}}
	\put(-120,100){$(c)$}	
	\caption{Trajectories by applying the policies trained in section \ref{Results_train_test} based on geometric symmetry in FRM. Note that the polices trained in the ${\romannumeral 3}$ sub-quadrant (see the solid lines) are applied directly to other sub-quadrants (the dashed lines) without new training. ($a$) $h_0=0.05\sqrt{2}$ and $\alpha_0=60^{\circ},80^{\circ}$. ($b$) $h_0=0.1\sqrt{2}$ and $\alpha_0=60^{\circ},80^{\circ}$. ($c$) $h_0=0.25\sqrt{2}$ and $\alpha_0=45^{\circ}$. }
	\label{Traj_sym}
\end{figure}

For instance, in figure \ref{FRM_act}$(a)$, the red droplet in the sub-quadrant ${\romannumeral 4}$ is symmetric to the blue droplet in the sub-quadrant ${\romannumeral 3}$ about the antidiagonal line ($-45^o$). In this case, the state of the red droplet can be expressed in terms of that of the blue droplet by $s^{\romannumeral 4}(t)=[-y(t),\ -x(t),\ -v(t),\ -u(t),\ -k_y(t),\ -k_x(t)]$. The relation can also be written in a matrix form \DX{(with $s^{\romannumeral 4}(t), s^{\romannumeral 3}(t)$ interpreted as columns) 
\begin{align}
s^{\romannumeral 4}(t)=\underbrace{\begin{pmatrix}
0 & -1 & 0 & 0 & 0 & 0 \\
-1 & 0 & 0 & 0 & 0 & 0 \\
0 & 0 & 0 & -1 & 0 & 0 \\
0 & 0 & -1 & 0 & 0 & 0 \\
0 & 0 & 0 & 0 & 0 & -1 \\
0 & 0 & 0 & 0 & -1 & 0 \\
\end{pmatrix}}_{L_g\text{ for antidiagonal line } (-45^\circ)}
s^{\romannumeral 3}(t)
\end{align}}
\DX{The $L_g$'s utilising the symmetry with respect to the horizontal axis, vertical axis and diagonal line read respectively
\begin{align}
\underbrace{\begin{pmatrix}
1 & 0 & 0 & 0 & 0 & 0 \\
0 & -1 & 0 & 0 & 0 & 0 \\
0 & 0 & 1 & 0 & 0 & 0 \\
0 & 0 & 0 & -1 & 0 & 0 \\
0 & 0 & 0 & 0 & 1 & 0 \\
0 & 0 & 0 & 0 & 0 & -1 \\
\end{pmatrix}}_{L_g \text{ for horizontal line}}, \quad
\underbrace{\begin{pmatrix}
-1 & 0 & 0 & 0 & 0 & 0 \\
0 & 1 & 0 & 0 & 0 & 0 \\
0 & 0 & -1 & 0 & 0 & 0 \\
0 & 0 & 0 & 1 & 0 & 0 \\
0 & 0 & 0 & 0 & -1 & 0 \\
0 & 0 & 0 & 0 & 0 & 1 \\
\end{pmatrix}}_{L_g \text{ for vertical line}}\quad \text{and} \quad
\underbrace{\begin{pmatrix}
0 & 1 & 0 & 0 & 0 & 0 \\
1 & 0 & 0 & 0 & 0 & 0 \\
0 & 0 & 0 & 1 & 0 & 0 \\
0 & 0 & 1 & 0 & 0 & 0 \\
0 & 0 & 0 & 0 & 0 & 1 \\
0 & 0 & 0 & 0 & 1 & 0 \\
\end{pmatrix}}_{L_g \text{ for diagonal line}\ (45^\circ)}.
\end{align}
Note that the diagonal line in this work is defined as the one with an angle of $45^\circ$.  } 

To apply the policy $\pi_\theta$ already trained in the sub-quadrant ${\romannumeral3}$ to ${\romannumeral4}$, we also need to consider the action in the sub-quadrant $iii$ to be transformed to \DX{${\omega}^{\romannumeral4}(t)=[-\Omega,\ +a_1(t)\Omega,\ -a_2(t)\Omega,\ +\Omega]$ } according to the symmetry with respect to the antidiagonal line, \DX{or ${\omega}^{\romannumeral4}(t)=K_g^s{\omega}^{\romannumeral3}(t)$,
\begin{align}
{\omega}^{\romannumeral4}(t)=\underbrace{\begin{pmatrix}
0 & 0 & -1 & 0 \\
0 & -1 & 0 & 0 \\
-1 & 0 & 0 & 0 \\
0 & 0 & 0 & -1 \\
\end{pmatrix}}_{K_g^s \text{ for antidiagonal line }  (-45^\circ)}
{\omega}^{\romannumeral3}(t).
\end{align}}
In this case, the rotation directions of all rollers are reversed as indicated by the red arrows in panel ($a$). Similarly, figure \ref{FRM_act}($b$) shows that the symmetry with respect to the vertical axis can be leveraged to control the droplet initially positioned in the sub-quadrants $ii$ based on the control policy trained in sub-quadrants $iii$. \DX{The $K_g^s$'s utilising the symmetry with respect to the horizontal axis, vertical axis and diagonal line read respectively
\begin{align}
\underbrace{\begin{pmatrix}
0 & 0 & 0 & -1 \\
0 & 0 & -1 & 0 \\
0 & -1 & 0 & 0 \\
-1 & 0 & 0 & 0 \\
\end{pmatrix}}_{K_g^s \text{ for horizontal line}}, \quad
\underbrace{\begin{pmatrix}
0 & -1 & 0 & 0 \\
-1 & 0 & 0 & 0 \\
0 & 0 & 0 & -1 \\
0 & 0 & -1 & 0 \\
\end{pmatrix}}_{K_g^s \text{ for vertical line}} \quad \text{ and } \quad
\underbrace{\begin{pmatrix}
-1 & 0 & 0 & 0 \\
0 & 0 & 0 & -1 \\
0 & 0 & -1 & 0 \\
0 & -1 & 0 & 0 \\
\end{pmatrix}}_{K_g^s \text{ for diagonal line }  (45^\circ)}.
\end{align}
}
The idea can be further extended to the remaining sub-quadrants. To sum up, the corresponding states and actions can be summarised as
\begin{itemize}
	\item \noindent {\romannumeral1}: $s^{\romannumeral 1}(t)=[y(t),\ -x(t),\ v(t),\ -u(t),\ k_{y}(t),\ -k_{x}(t)]$; ${\omega}^{\romannumeral1}(t)=[-a_1(t)\Omega,\ +\Omega,\ -\Omega,\ +a_2(t)\Omega]$;
	\item \noindent{\romannumeral2}: $s^{\romannumeral 2}(t)=[-x(t),\ y(t),\ -u(t),\ v(t),\ -k_{x}(t),\ k_{y}(t)]$; ${\omega}^{\romannumeral2}(t)=[+a_1(t)\Omega,\ -a_2(t)\Omega,\ +\Omega,\ -\Omega]$;
	\item \noindent{\romannumeral 5}: $s^{\romannumeral 5}(t)=[-y(t),\ x(t),\ -v(t),\ u(t),\ -k_{y}(t),\ k_{x}(t)]$; ${\omega}^{\romannumeral5}(t)=[-\Omega,\ +a_2(t)\Omega,\ -a_1(t)\Omega,\ +\Omega]$;
	\item \noindent{\romannumeral6}: $s^{\romannumeral 6}(t)=[x(t),\ -y(t),\ u(t),\ -v(t),\ k_{x}(t),\ -k_{y}(t)]$; ${\omega}^{\romannumeral6}(t)=[+\Omega,\ -\Omega,\ +a_1(t)\Omega,\ -a_2(t)\Omega]$;
	\item \noindent {\romannumeral7}: $s^{\romannumeral 7}(t)=[-x(t),\ -y(t),\ -u(t),\ -v(t),\ -k_{x}(t),\ -k_{y}(t)]$; ${\omega}^{\romannumeral7}(t)=[+\Omega,\ -\Omega,\ +a_2(t)\Omega,\ -a_1(t)\Omega]$;
	\item \noindent{\romannumeral8}: $s^{\romannumeral 8}(t)=[y(t),\ x(t),\ v(t),\ u(t),\ k_{y}(t),\ k_{x}(t)]$; ${\omega}^{\romannumeral8}(t)=[-a_2(t)\Omega,\ +\Omega,\ -\Omega,\ +a_1(t)\Omega]$.
\end{itemize} 
In order to validate the idea based on the geometric symmetry, we apply the policies obtained in Section \ref{Results_train_test} for the sub-quadrants $iii$ to the entire flow domain directly without new training. As illustrated in Figure \ref{Traj_sym}, the dashed lines represent the direct application of the trained policies described in Section \ref{Results_train_test}. It is evident that all the droplets are successfully guided to the origin.

%

\subsection{\DX{The effect of inertia on the DRL control of FRM}}\label{Nonlinear_Inertia}
In this study, inertia is numerically modelled, and its effects on the DRL control will be discussed in this section. The $Re$ investigated ranges from $Re = 10^{-9}, 0.4, 2$ to $3$. One may wonder why we limit the study to relatively small values of $Re$. The reason is that, as $Re$ increases, control becomes progressively more challenging. At higher $Re$, the delay in flow response caused by the increased inertia disrupts the action-reward relationship in DRL, ultimately leading to failed control. This will be explained next.

\begin{figure}
	\centering
	\subfigure{\includegraphics[width=0.45\textwidth]{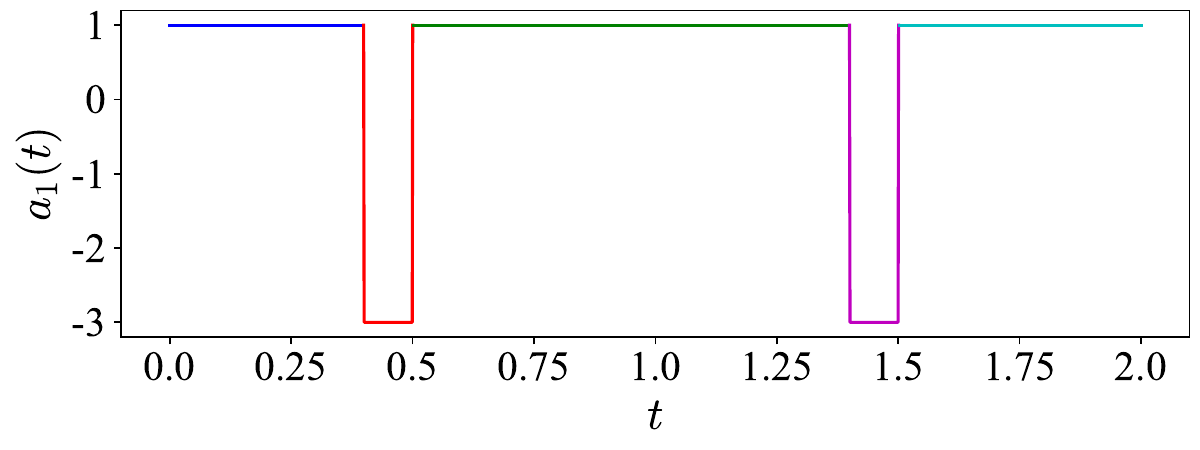}}	
	\put(-170,65){$(a)$}		
	\subfigure{\includegraphics[width=0.45\textwidth]{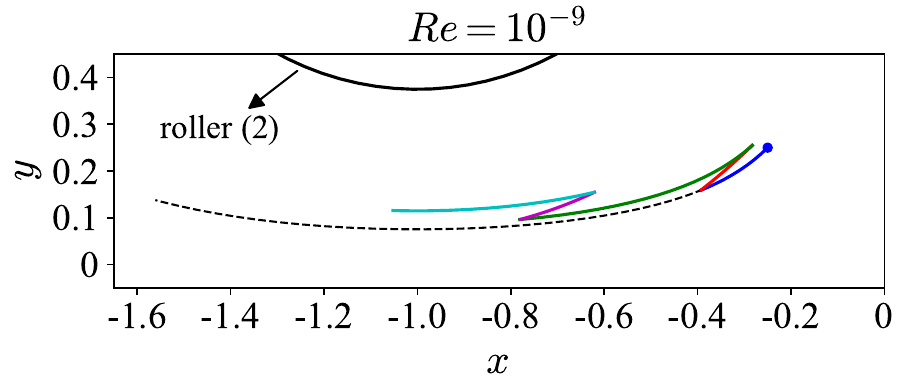}}
	\put(-170,65){$(b)$}\\
	\subfigure{\includegraphics[width=0.45\textwidth]{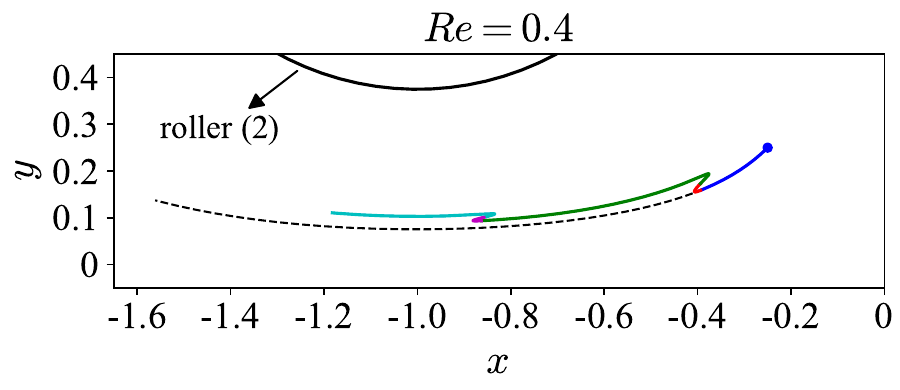}}
	\put(-170,65){$(c)$}
	\subfigure{\includegraphics[width=0.45\textwidth]{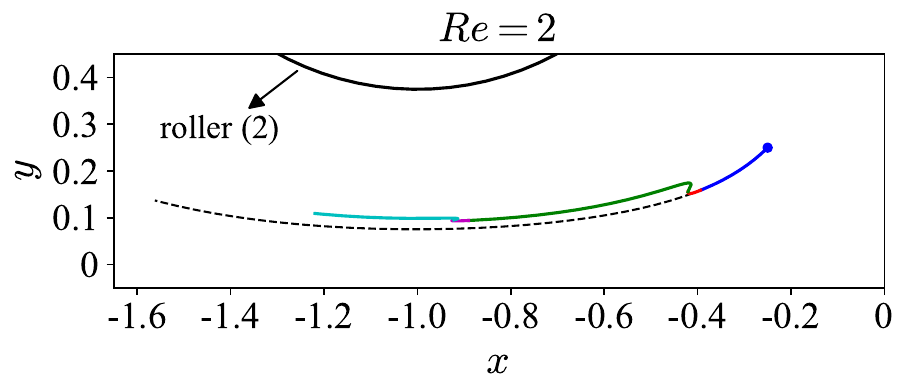}}
	\put(-170,65){$(d)$}\\
	\subfigure{\includegraphics[width=0.45\textwidth]{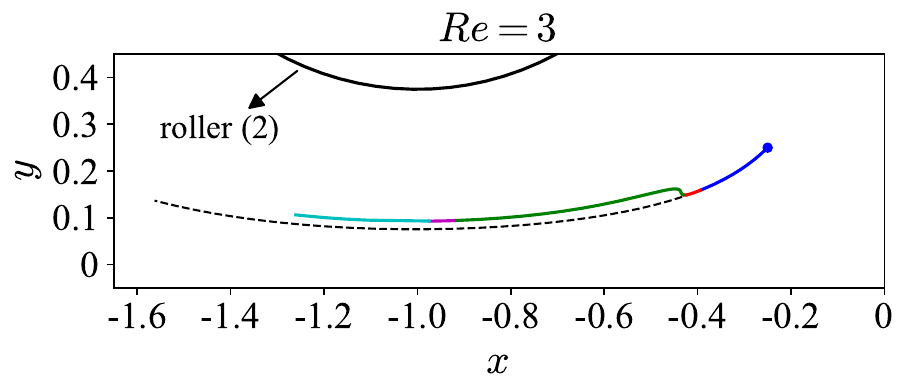}}
	 \put(-170,65){$(e)$}	
	\subfigure{\includegraphics[width=0.45\textwidth]{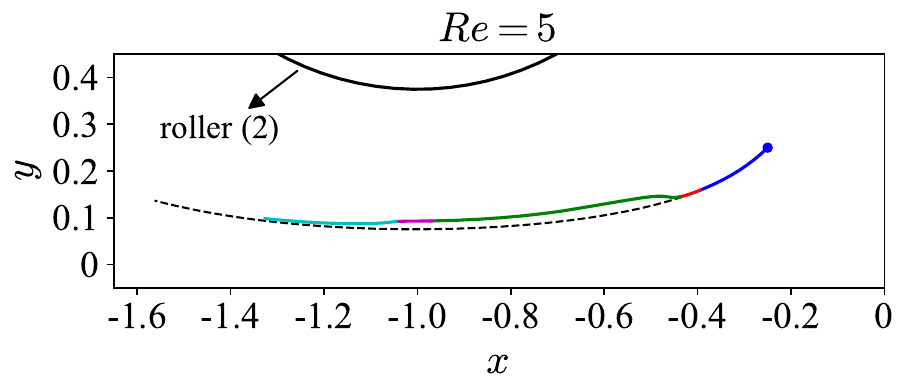}}
	 \put(-170,65){$(f)$}
	\caption{Effect of inertia on the droplet trajectory subject to \textit{ad hoc} action variation. The blue dot marks the initial position of the droplet at $[-0.25, 0.25]$. Panel ($a$) illustrates the action history of the roller (labeled as roller 2), where different colours represent the variations in the applied control action over time. In other panels, the black dashed lines show the trajectories without control, corresponding to the case where $a_1(t) = 1$ for $t \in [0, 2]$. The curve segments in multiple colours show the trajectories of the droplet under the corresponding actions. }
	\label{Impulse_traj}
\end{figure}

Figure \ref{Impulse_traj} illustrates the effect of inertia on the delay in flow response by examining the droplet trajectory as the applied action is varied. This demonstration is an \textit{ad hoc} test and does not correspond to the control cases in table \ref{Train_parameters}. In panels ($b$-$f$), the black dashed line represents the trajectory of a droplet started at the initial position $[-0.25, 0.25]$ without control, following the baseline roller action $[+\Omega, -\Omega, +\Omega, -\Omega]$. Since the $Re$'s are small, all these black dashed trajectories appear similar, although minor differences exist that are difficult to discern. However, when the roller action is varied to $[+\Omega, -a_1(t)\Omega, +\Omega, -\Omega]$, with the profile of $a_1(t)$ shown in panel $(a)$, the variation in the trajectories across the considered $Re$'s becomes significant, even though the values of $Re$ are generally small. This is depicted by the curves in multiple colors. Notably, for $Re = 10^{-9}$, where inertia is negligible, the droplet instantly adjusts its motion in response to changes in the roller action. As $Re$ increases, the effects of inertia become more pronounced. The droplet exhibits greater resistance to changes in direction, despite the roller action being reversed with a large amplitude. This is particularly evident in high-$Re$ cases, and leads to significant delay in flow response, potentially disrupting the action-reward relationship in the DRL control. For example, at $Re=3$ and $Re=5$, the red segments of the trajectories appear to follow the continuation of the blue segments, even though the red signal represents a reversed rotation with relatively large amplitude. As the DRL agent interprets the outcome of each control action and refines the policy based on the perceived flow state, it may mistakenly infer that the red action has no influence on the droplet’s position. This misinterpretation can ultimately lead to a failed control attempt if the state space is defined solely based on position. In our DRL setup, the state includes position, velocity, and acceleration. However, even with this comprehensive state representation, the current approach fails to achieve successful control for $Re=5$ in our FRM flow. Consequently, results for $Re=5$ are not included in this manuscript.

In the following, we will provide the control performance of the DRL agent for $Re=10^{-9}$ and $Re=2,3$, to be compared with the results in Sec. \ref{Results_train_test} for $Re=0.4$.

\subsubsection{Vanishingly-small $Re$}\label{vanishingRe}
\begin{figure}
	\centering
	\subfigure{\includegraphics[height=0.31\textwidth,trim= 0 10 0 5,clip]{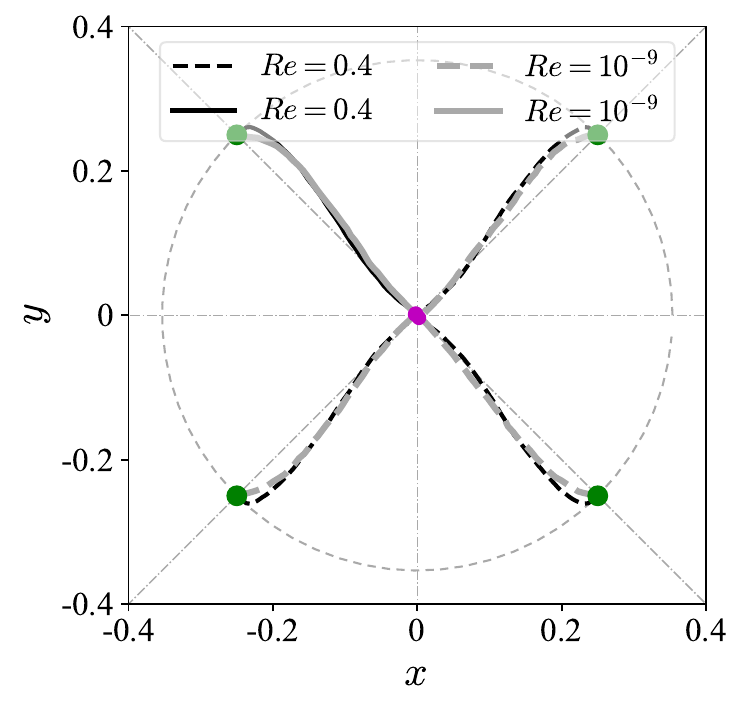}}	
	\put(-130,110){$(a)$}
	\subfigure{\includegraphics[height=0.31\textwidth,trim= 0 10 0 5,clip]{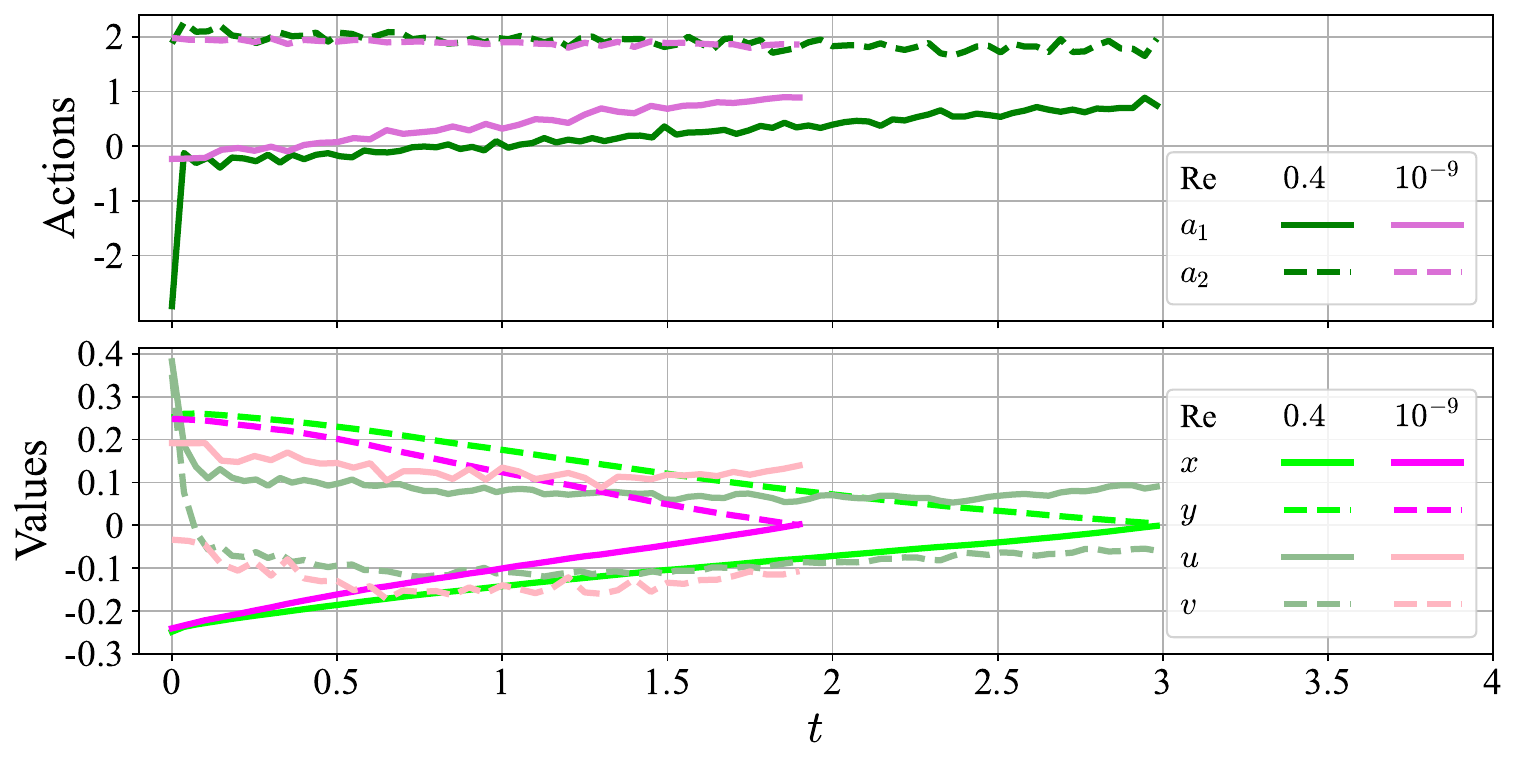}}
	\put(-250,110){$(b)$}
	\caption{Droplet trajectory under control corresponding to the case {\color{black}5.1} in table \ref{Train_parameters} with the parameters $h_0 = 0.25\sqrt{2}$, $\alpha_0 = 45^{\circ}$ and $Re=10^{-9}$, compared to the baseline case {\color{black}5.2} ($Re=0.4$), which has been explained in detail in section \ref{Results_train_test}. ($b$) The roller actions in the case of $h_0 = 0.25\sqrt{2}$ and $\alpha_0 = 45^{\circ}$ and the value history of the positions $[x{(t)},\ y{(t)}]$ and velocities $[u{(t)},\ v{(t)}]$.}
	\label{fig:vanishingRe}
\end{figure}

For completeness, we report control results for a vanishingly-small-$Re$ ($=10^{-9}$) flow in FRM to elucidate the differences in the numerical settings and results between the vanishingly-small-$Re$ and the finite $Re=0.4$ cases. Figure \ref{fig:vanishingRe}$(a)$ shows that the {\color{black} furthest} case $h_0=0.25\sqrt{2}, \alpha_0=45^{\circ}$ in the small-$Re$ flow can be controlled successfully by a DRL agent trained with only the droplet position as the state, without needing velocity or acceleration. In contrast, in the finite-$Re$ cases, the converged DRL agent entails \DX{velocity component}, highlighting its increased complexity compared to the small-$Re$ flow. \DX{The test presented in Appendix \ref{statedef} suggests that acceleration may play a less significant role compared to velocity in defining the state. } The figure also {\color{black} demonstrates} that the inertial effects result in more curved control trajectories than in the Stokes case, \DX{consistent with our \textit{ad hoc} test in figure \ref{Impulse_traj}}. The geometric symmetry has been leveraged in panel \ref{fig:vanishingRe}($a$). Compared to the small-$Re$ case in \cite{vona2021stabilizing}, where droplet trajectories exhibit zigzags, our \DX{trajectories display less wiggles}. Figure \ref{fig:vanishingRe}$(b)$ compares the actions and the positions/velocities of the two flows. Two notable differences between them can be observed: (1) the Stokes flow can be controlled in a shorter time, and (2) the actions in the \DX{$Re=0.4$ } case deviate more from the baseline rotation compared to those in the Stokes flow. These differences again highlight the increased complexity brought about by the inertia in the finite-$Re$ case.

\subsubsection{\DX{Training results for $Re=2,3$}}\label{result_Re3}

\begin{figure}
	\centering
	\includegraphics[width=0.33\textwidth]{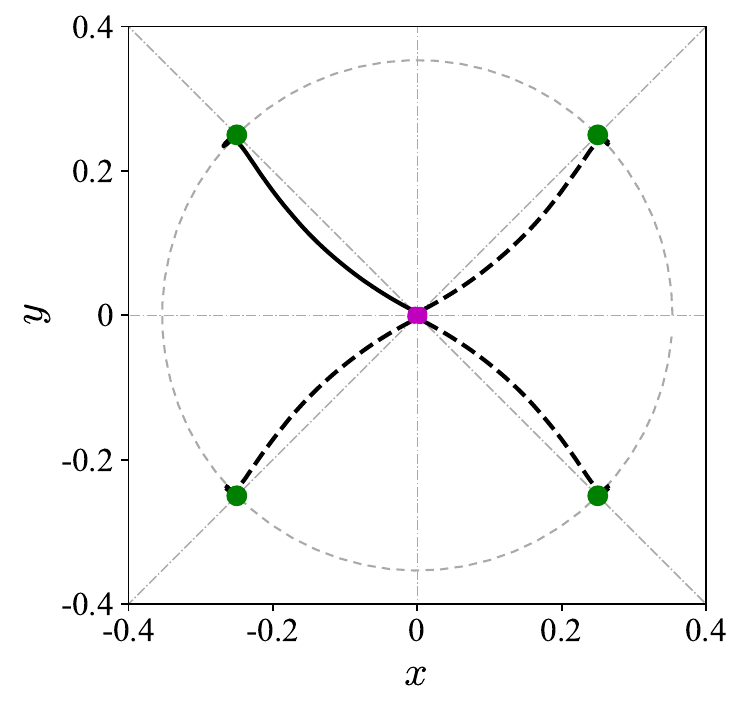}
	\put(-130,115){$(a)$}	
	\includegraphics[width=0.62\textwidth]{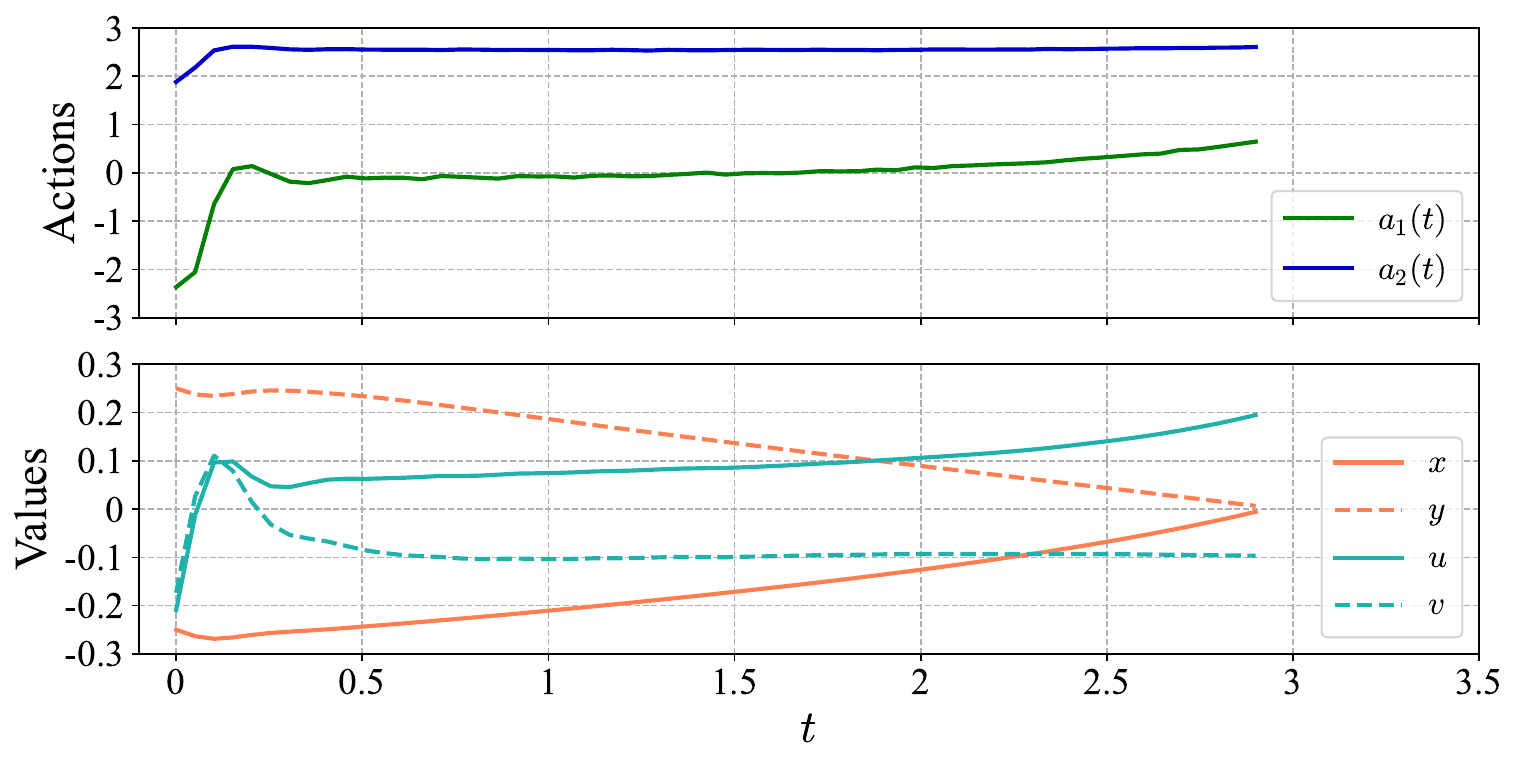}
	\put(-240,115){$(b)$}	\\
	\includegraphics[width=0.33\textwidth]{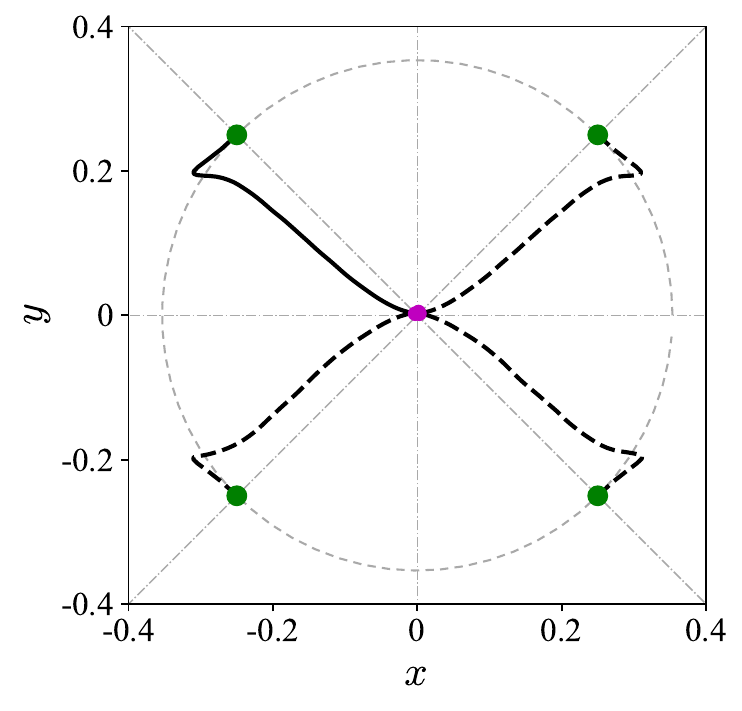}
	\put(-130,115){$(c)$}	
	\includegraphics[width=0.62\textwidth]{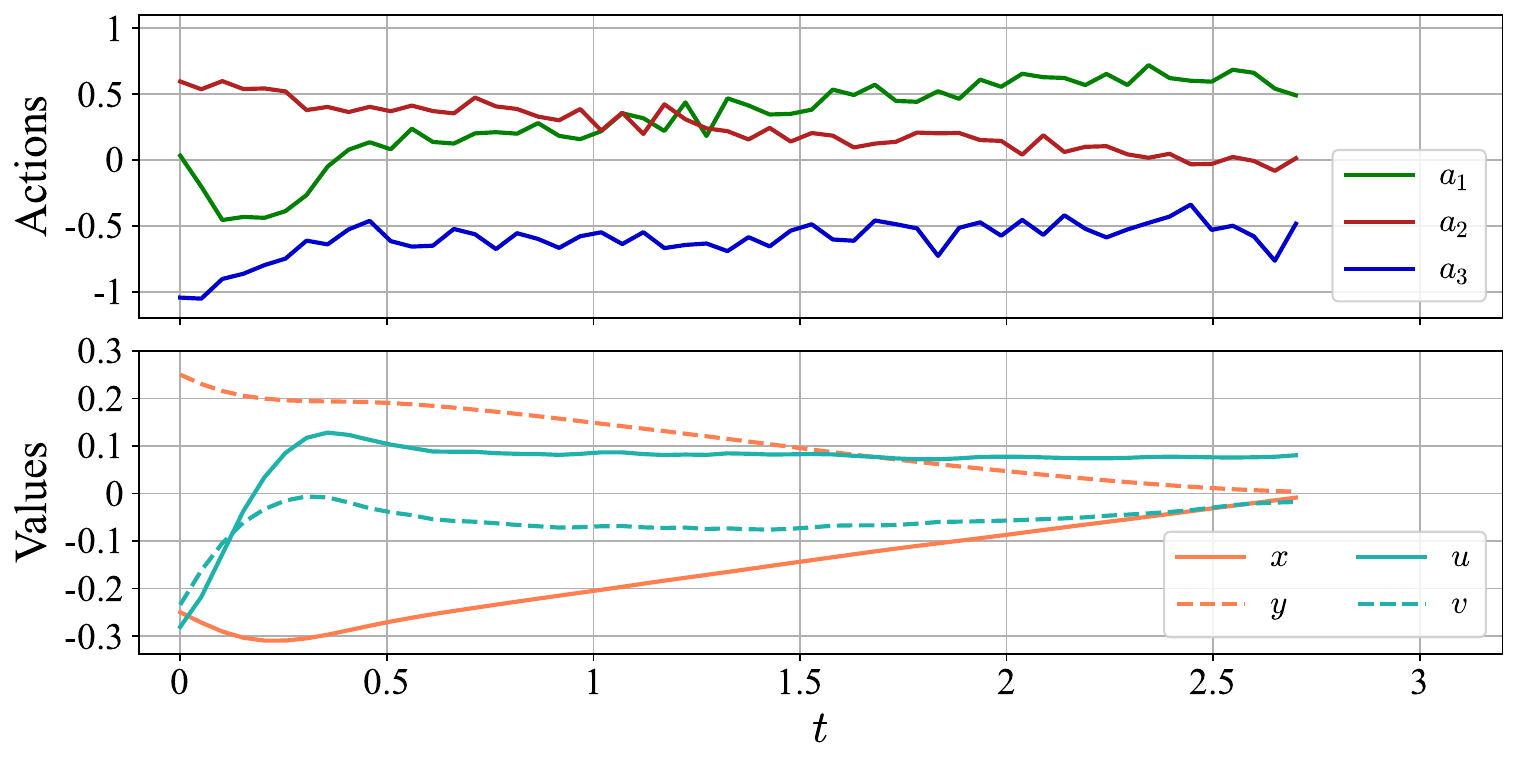}
	\put(-240,115){$(d)$}
	\caption{($a$) Controlled droplet trajectory for case {\color{black}5.3} in Table \ref{Train_parameters} with the parameters $h_0 = 0.25\sqrt{2}$, $\alpha_0 = 45^{\circ}$ and $Re=2$. ($b$) Action history and value history in the test. ($c$) Controlled droplet trajectory for case {\color{black}5.4} in Table \ref{Train_parameters} with the parameters $h_0 = 0.25\sqrt{2}$, $\alpha_0 = 45^{\circ}$ and $Re=3$. ($d$) Action history and value history in the test. Three rollers are activated in this control task.}
	\label{Traj_Re3}
\end{figure}

This section presents additional results for \(Re = 2\) and \(Re = 3\), as shown in Figure \ref{Traj_Re3}. For \(Re = 3\), the increasing inertia results in a more significant delay in flow response. Our numerical tests indicate that introducing an additional roller is necessary for effective control in this case. Specifically, rollers (1), (2), and (3) are utilized, corresponding to actions \(a_1(t)\), \(a_2(t)\), and \(a_3(t)\), respectively. Figure \ref{Traj_Re3} panels ($a$) and ($b$) illustrate successful training outcomes for \(Re = 2\). Similar to the \(Re = 0.4\) case, the action \(a_2(t)\) in this case takes on positive values, while \(a_1(t)\) initially assumes negative values before transitioning to positive. The controlled trajectory exhibits a slight turn at the beginning, which becomes more pronounced for \(Re = 3\), as shown in panel ($c$). The trajectory for \(Re = 3\) suggests that a strong force pointing towards bottom-left is exerted on the droplet by the extensional flow. The collective actions of the rollers successfully guide the droplet toward the target. The more distorted trajectory pattern observed for \(Re = 3\), combined with the need for three rollers, highlights the increased difficulty of control with higher flow inertia. This challenge is closely linked to the delayed flow response, as discussed earlier in this section. Future research could focus on enhancing the controllability of DRL systems for higher-\(Re\) flows by leveraging more advanced algorithms and control strategies.

\section{{Conclusion}}
In this study, we have further tested the applicability of DRL in controlling droplet trajectories within a FRM simulated by DNS. The work extends that of \cite{bentley1986computer,vona2021stabilizing} but with two new considerations. First, we focused on the finite-$Re$ regime, incorporating nonlinear inertial effects in our control problem. Second, we have leveraged the geometric symmetry of the FRM to enhance the training efficiency of the DRL policies. 

Our results have demonstrated that DRL can effectively harness the complex flow dynamics of FRM to achieve desired droplets movement towards the origin, even when starting from challenging initial conditions that place the droplets far from the center. The ability of the DRL agent to adaptively adjust \DX{two or even three } roller speeds demonstrates its effectiveness. The effect of inertia in the control task has also been discussed from the perspective of flow physics. \DX{We note that the delay in flow response caused by the inertial effect can potentially disrupt the action-reward relationship in DRL, which makes the precise control more challenging as the $Re$ increases. Future efforts are needed to improve the performance of DRL controllers in FRM flows with higher $Re$.}

In addition, the investigation into the intrinsic symmetry of FRM has provided a better practice that enables the application of trained control policies across various sub-quadrants of the flow field without loss of performance. This idea can be readily applied to other control methods for FRM. 

\DX{Future work could focus on further optimisation of control in high-$Re$ regimes, where inertia becomes more dominant. Additionally, incorporating more advanced noise-handling techniques or extending the DRL framework to multi-agent settings could improve the performance of RL in dealing with multiple initial conditions. More refined reward functions can further enhance the control performance, such as a penalty term to avoid abrupt changes in angular velocities } {\color{black} and torque fluctuation}. Other future research includes extending the DRL strategy to handle more complex flows in FRM. For example, polymeric flows at vanishingly-small $Re$ may exhibit elastic nonlinearity. Our attempt to control a nonlinear flow with finite $Re$ may showcase the applicability of DRL in these complex fluids. 

Declaration of Interests. The authors report no conflict of interest.

M.Z. acknowledges the financial support of a Tier 1 grant from the Ministry of Education, Singapore (WBS No. A-8001172-00-00).

\appendix
\section{{\DX{Action updating time interval $\Delta t_a$}}}\label{appendix_ta}
\begin{figure}
	\centering
	{\includegraphics[width=0.49\textwidth]{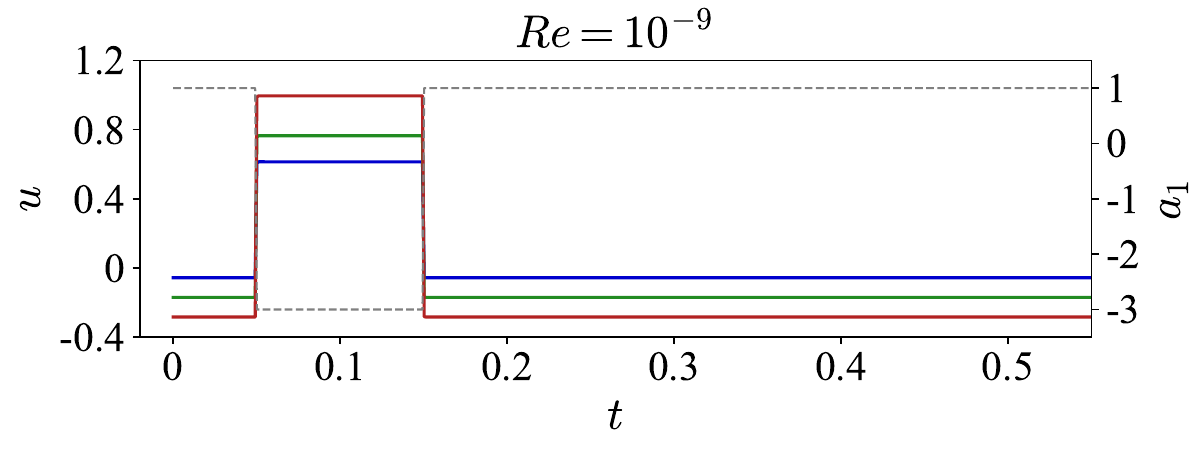}}
	\put(-192,60){$(a)$}
	{\includegraphics[width=0.49\textwidth]{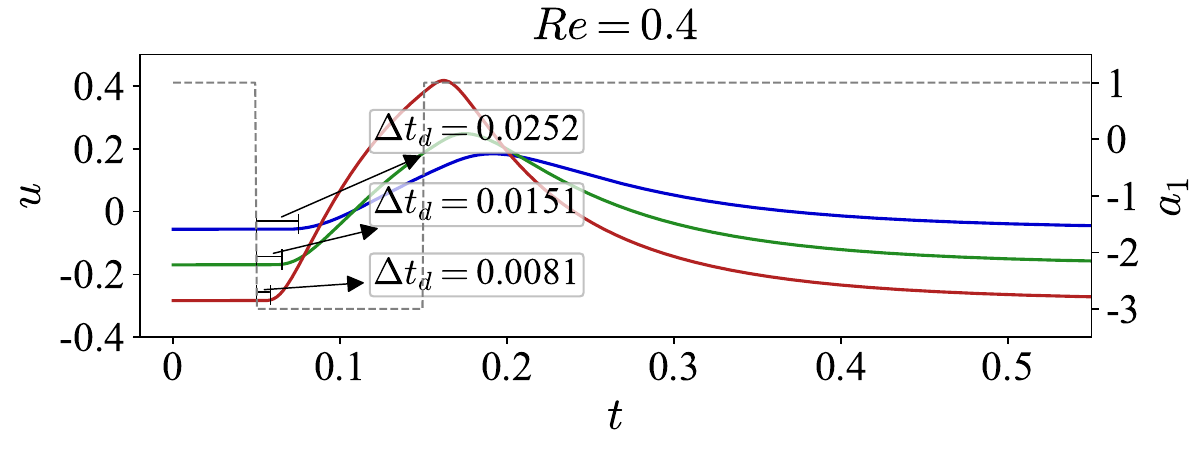}}
	\put(-192,60){$(b)$}\\
	{\includegraphics[width=0.49\textwidth]{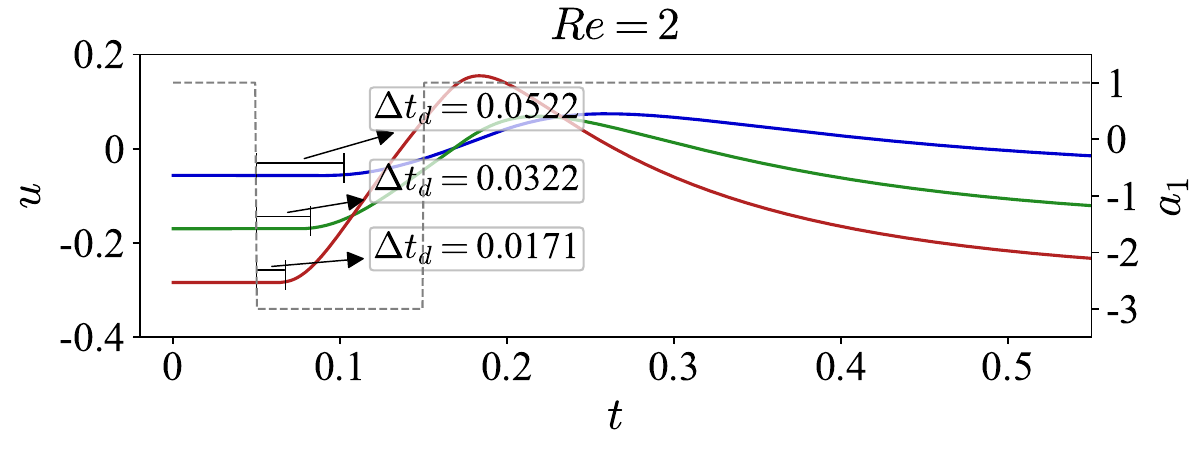}}
	\put(-192,60){$(c)$}
	{\includegraphics[width=0.49\textwidth]{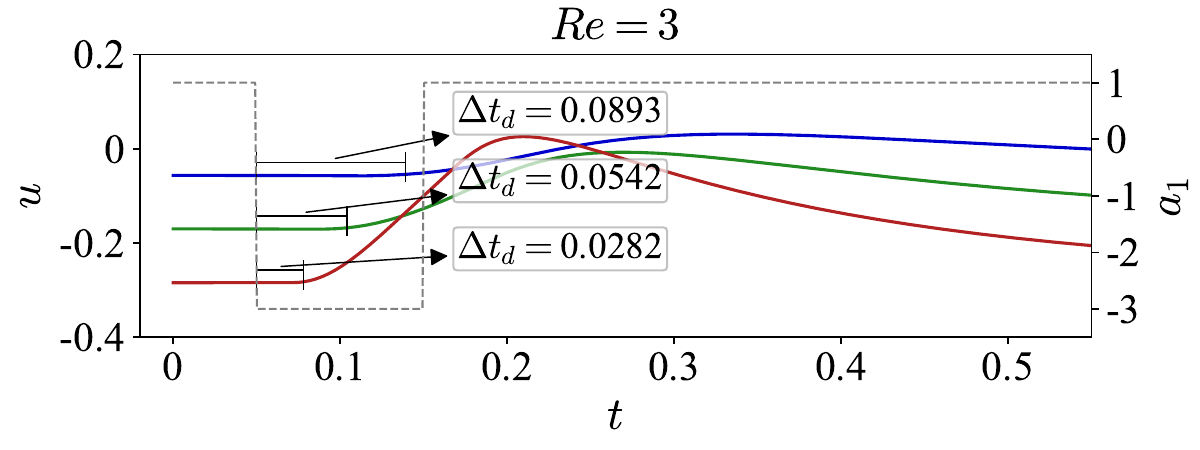}}
	\put(-192,60){$(d)$}
	\caption{The response test across $Re$'s. The dashed line represents the actions exerted by roller (2), denoted as $a_1(t)$. The red, green and blue lines are the velocity signals monitored at positions of $h_0=[0.25\sqrt{2},0.05\sqrt{2},0.25\sqrt{2}]$ and $\alpha_0=45^{\circ}$. }
	\label{Impulse_response_app}
\end{figure}

This appendix outlines a heuristic approach to determine an appropriate action update time interval, $\Delta t_a$. This quantity is closely related to the delay in flow response, as discussed in Sec.~\ref{Nonlinear_Inertia}. To quantitatively determine $\Delta t_a$, we estimate the response time $\Delta t_d$, defined as the duration between the initiation of an action and the onset of a $0.1\%$ velocity change at a certain position. Figure~\ref{Impulse_response_app} presents an \textit{ad hoc} test of the flow response time $\Delta t_d$ for various $Re$. In this test, the action is varied as shown by the thin dashed lines (corresponding to the right $y$-axis for $a_1(t)$). Specifically, the roller's action changes from the default value $1$ to $-3$ over the interval $t = [0.05, 0.15]$ before returning to its default value. Three velocity probes are placed at positions $[0.05\sqrt{2}, 0.15\sqrt{2}, 0.25\sqrt{2}]$ with $\alpha_0 = 45^\circ$ to monitor the corresponding flow responses.

The estimated $\Delta t_d$ values are reported in the figure. Our experiments show that, for successful control policy training, the action update time interval $\Delta t_a$ should generally be of the same order as, or larger than, the flow response time $\Delta t_d$. This ensures the agent has sufficient time to evaluate the consequences of its actions within the flow environment. Based on these observations, we set $\Delta t_a = 0.05$ for $Re = [0.4, 2]$. For $Re = 3$, the flow response time is notably larger, as seen in Fig.~\ref{Impulse_response_app}(d). Accordingly, the action is updated every 75 numerical time steps, corresponding to $\Delta t_a = 0.075$. It is worth noting that these values of $\Delta t_a$ are not necessarily optimal, and there remains room for further improvement.  

\section{\DX{Robustness test with thermal noise}}\label{result_noise}

We have established effective control policies in deterministic flow environment. In this appendix, we test the trained policy under thermal noise. Based on the work by \cite{vona2021stabilizing}, we implement the Langevin approach by adding the thermal noise term to the NS equation which can be expressed as
\begin{equation}
	F_i = (2 \Delta t / \mathrm{Pe})^{1 / 2} \Gamma_i
	\label{Eq_thermal}
\end{equation}
where ${\rm Pe}$ is the dimensionless P\'eclet number defining the relative magnitude of the advection over Brownian diffusion. \mbox{Eq.\ \ref{Eq_thermal}} is added at the end of each numerical step, where $\Delta t$ is the numerical time interval and $\Gamma_i$ (with $i = u, v$) is drawn from a standard normal distribution.

A lower ${\rm Pe}$ number results in higher thermal noise, which can interfere with the control. The following experiment sets a range of ${\rm Pe}$ numbers and tests the policies to examine whether they can still guide the droplets back to the origin within the target distance. The test uses the policies for the farthest initial positions, namely $h_0 = 0.25\sqrt{2}$ and $\alpha_0 = 45^{\circ}$, for all the $Re$ considered. A control is considered successful if the final distance is less than the target value $h_e= 0.005$. 

\begin{figure}
	\centering
	\includegraphics[width=0.4\textwidth]{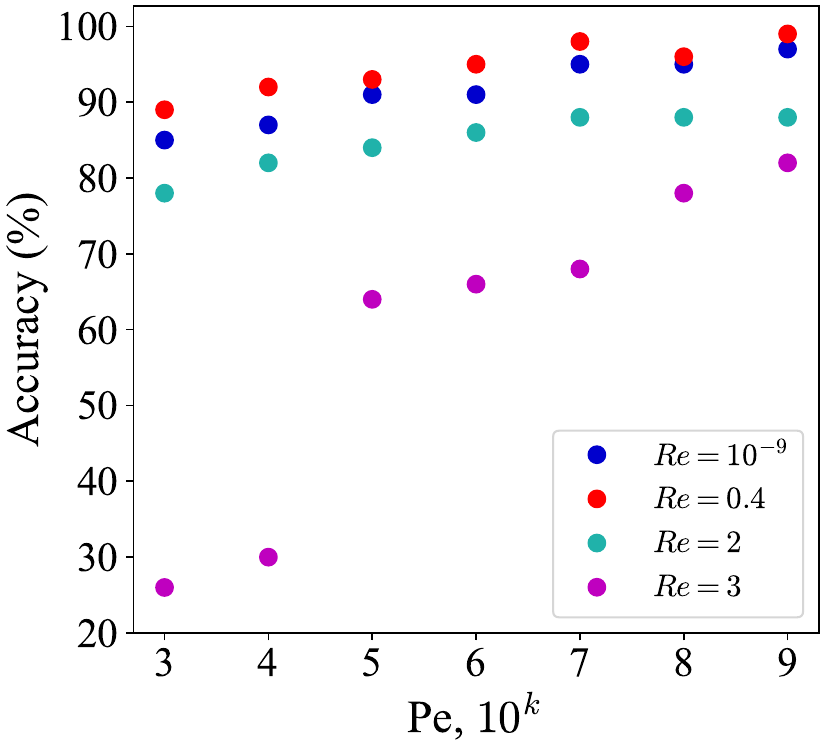}
	\put(-160,130){$(a)$}
	\quad\quad
	\includegraphics[width=0.4\textwidth]{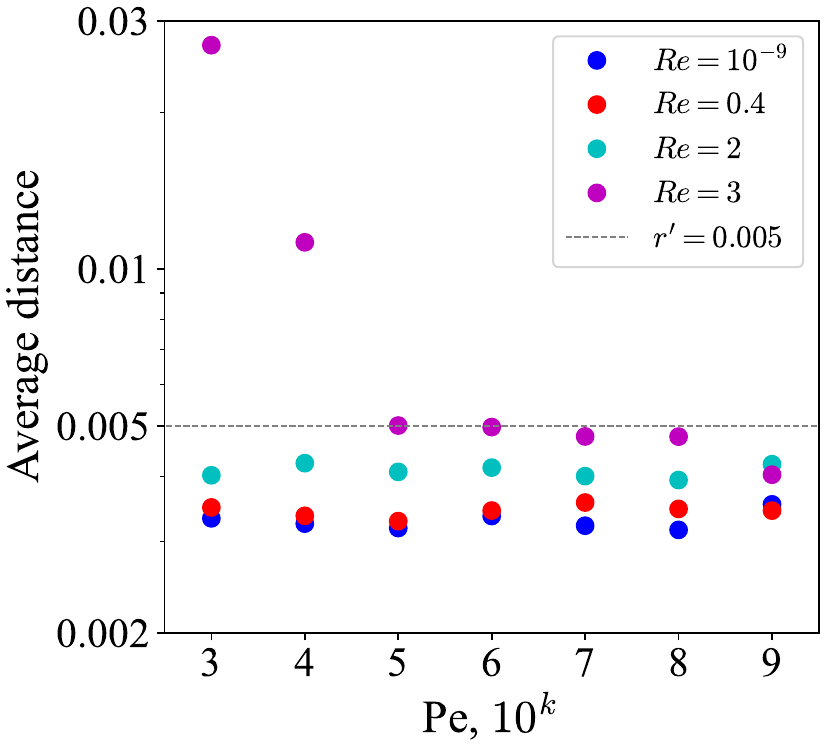}
	\put(-160,130){$(b)$}
	\caption{Robustness test based on adding thermal noise to the flow environment. (a) The accuracy as the number of successful controls out of 100 total runs. (b) The average distance of 100 total runs.}
	\label{Robustness_test}
\end{figure}

\mbox{Fig.\ \ref{Robustness_test}} shows the test results, where panels (a) and (b) display the accuracy and average distance, respectively. The accuracy is computed as the number of successful controls out of 100 total runs. From panel (a), one can clearly see that for all the policies tested, the accuracy decreases with smaller ${\rm Pe}$ numbers as noise levels increase. Among the Reynolds numbers, the greatest $Re$ (=3) is negatively influenced most by the thermal noise. This is understandable since the droplet's motion in this flow regime is most affected by inertia, making it the most complex case to control.

Panel (b) displays the average distance at the end of the testing step. In most cases (except for $Re=3$ at ${\rm Pe}=10^k, k=3,4$), the average distance is less than or approximately equal to the target distance $h_e = 0.005$. This demonstrates the effectiveness of the policies in a noisy environment.



\section{\DX{Global policy}}\label{result_nearbyICs}
\begin{figure}
	\centering
	\subfigure{\includegraphics[width=0.495\textwidth]{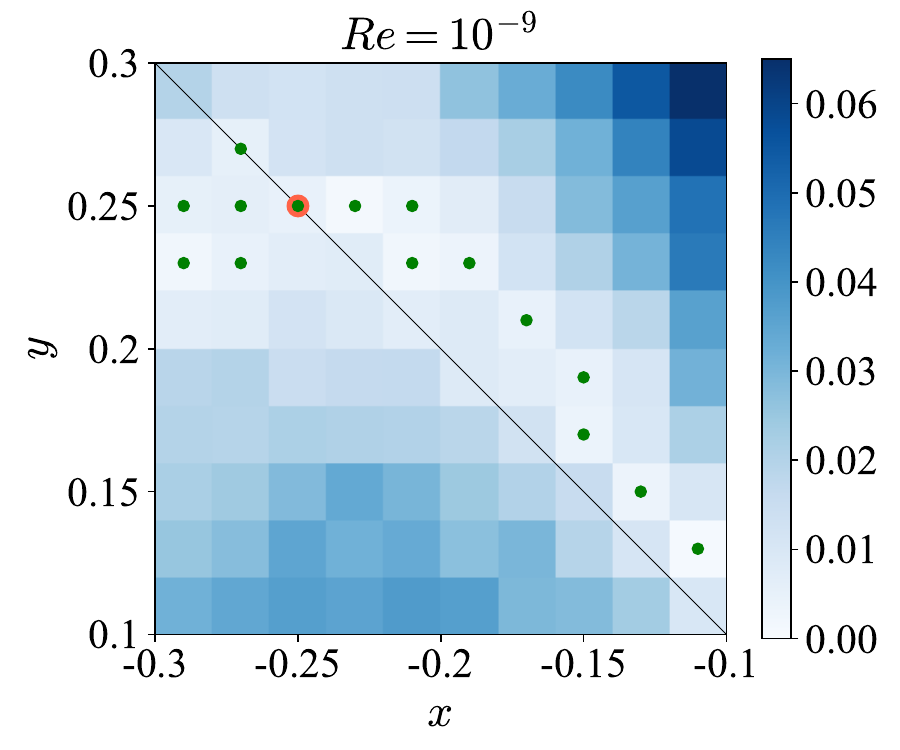}}
	\put(-170,130){$(a)$}	
	\subfigure{\includegraphics[width=0.495\textwidth]{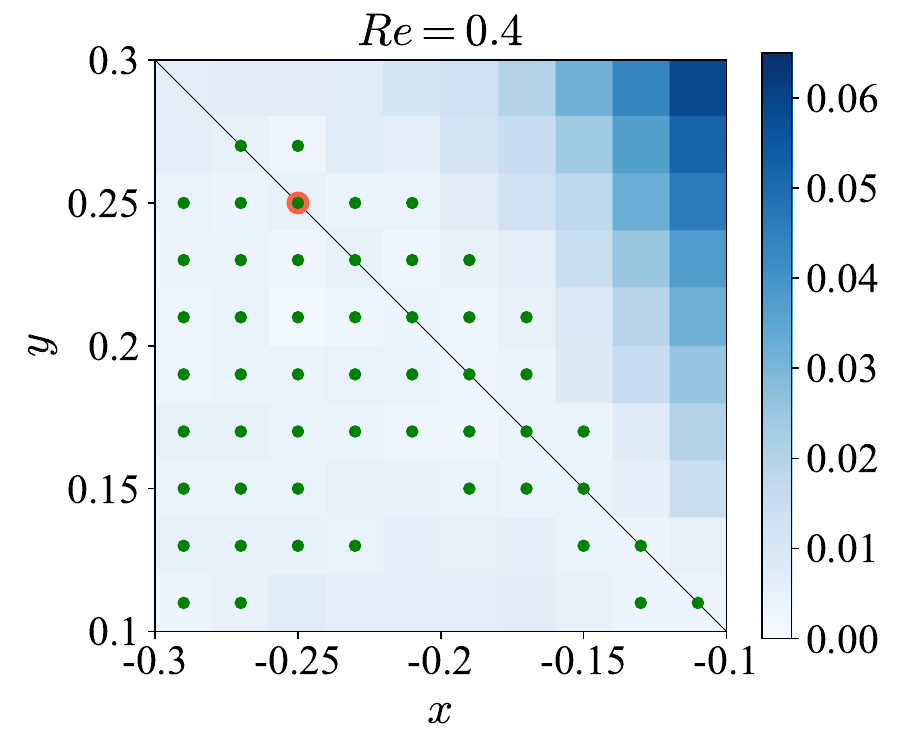}}
	\put(-170,130){$(b)$}	
	\newline
	\subfigure{\includegraphics[width=0.495\textwidth]{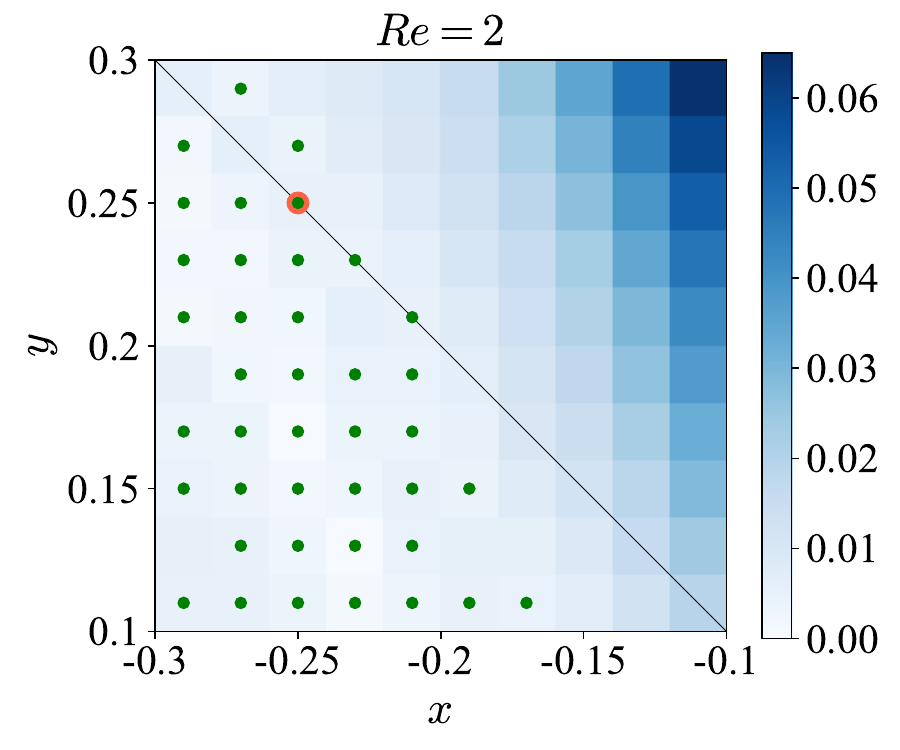}}
	\put(-170,130){$(c)$}	
	\subfigure{\includegraphics[width=0.495\textwidth]{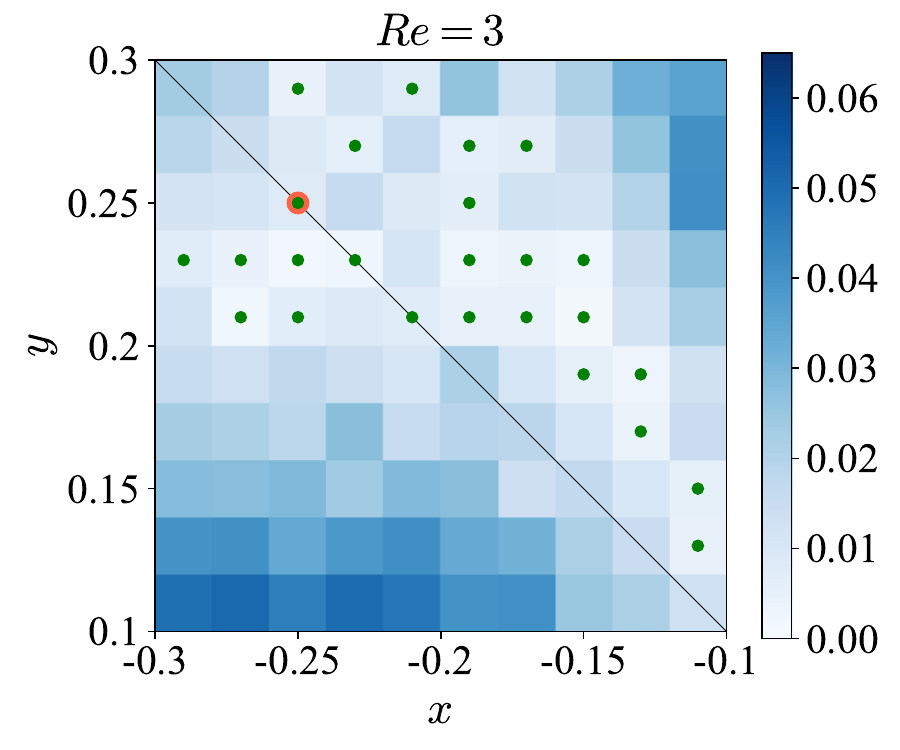}}
	\put(-170,130){$(d)$}	
	\caption{Final distances of droplets to the target position. Droplets are initially located at the cell centers on a $10 \times 10$ rectangular grid, uniformly distributed over $x \times y \in [-0.3, -0.1] \times [0.1, 0.3]$. The applied policy was trained for the specific initial position $(-0.25, 0.25)$, denoted by the red dot. The green dots indicate the final distance to the target is less than $h_e=0.005$. Some thermal noise with ${\rm Pe} = 10^{5}$ has been added to the flow environment in these tests.}
	\label{Nearby_test}
\end{figure}

\cite{vona2021stabilizing} investigated the possibility of a global policy by applying the policy trained for a specific case with $\bds{x}_0=(-0.03,0.02)$ to the other points a nearby region. The idea was to evaluate the performance of the trained policy generalised to different initial positions. 

To test the applicability of a similar global policy, we apply the trained policy for $h_0 = 0.25\sqrt{2}$ and $\alpha_0 = 45^{\circ}$ to other points within the region $x\times y \in [-0.3, -0.1] \times [0.1, 0.3]$. The region is divided into a \(10 \times 10\) rectangular grid of evenly spaced cells, as shown in Figure \ref{Nearby_test}. Trajectories are simulated starting from the center of each grid cell using the policies trained for the initial condition \((-0.25, 0.25)\), indicated by the red dot. The final distance to the origin is shown for each trajectory as a colour map in figure \ref{Nearby_test}. Some thermal noise with ${\rm Pe} = 10^{5}$ has been added to the flow environment. It is noted that our lattice is significantly greater than that considered in \cite{vona2021stabilizing} and the droplets are placed further away from the target.

Consistent with the findings in \cite{vona2021stabilizing}, figure \ref{Nearby_test} shows that points closer to the position $(-0.25, 0.25)$ tend to have smaller final distances to the target, particularly those near the diagonal line. Unsuccessful cases (regions without green dots) are predominantly clustered along the edges of the domain, likely due to the strong extensional flow in these areas. Some differences are observed across the considered $Re$'s. Specifically, for intermediate-$Re$ values, the lower-triangular region exhibits more successful testing outcomes.

{\color{black}In addition, we note that the trained policy, developed for a specific $Re$ and geometry (e.g., roller radius), can also be effectively applied to similar settings (results not shown). This suggests that the trained policy exhibits a degree of generalization, allowing it to perform well under conditions that are close to those for which it was originally trained.}

\section{\DX{Refining the hyperparameters in the reward function}}\label{hyper}

This appendix explains how the hyperparameters are determined in our reward function as in Eq. \ref{reward}. We will study  the effect of each term, i.e., $r_1(t),r_2(t), r'$ and the selected values of the parameters used therein. For clarity, we will focus on $[h_0,\alpha_0]=[0.05\sqrt{2},45^{\circ}]$ and $Re=0.4$ to illustrate.

\subsection{The ${r_1(t)}$ term} 
\begin{figure}
	\centering
	\subfigure{\includegraphics[width=0.4\textwidth]{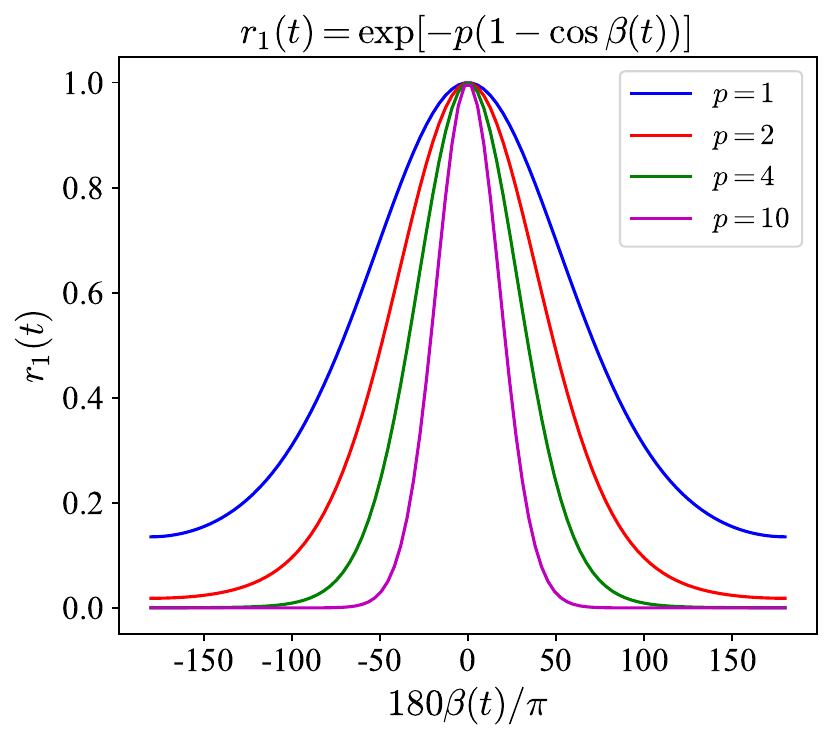}}
	\caption{The first reward term $r_1(t)$ against the angle in degrees between two displacement vector.}
	\label{r1_optimal}
\end{figure}

We discuss first the parameter $p$ which appears in the first term of the reward function, $r_1(t) = \exp[-p(1 - \cos \beta(t))]$, where $\beta(t)$ is the angle between two consecutive displacement vectors (see figure 1 in the manuscript). Figure \ref{r1_optimal} below plots $r_1(t)$ as a function of $\beta(t)$, showing that a larger $p$ value incentivizes the droplet to move in directions confined within a narrower angle between the displacement vectors. This suggests that $p$ plays a role in controlling the exploration-exploitation trade-off in the DRL control. Specifically, a smaller $p$ promotes exploration, though it may hinder policy convergence, while a larger $p$ encourages exploitation, sacrificing the exploration capacity.

\begin{figure}
	\centering
	\subfigure{\includegraphics[width=0.4\textwidth]{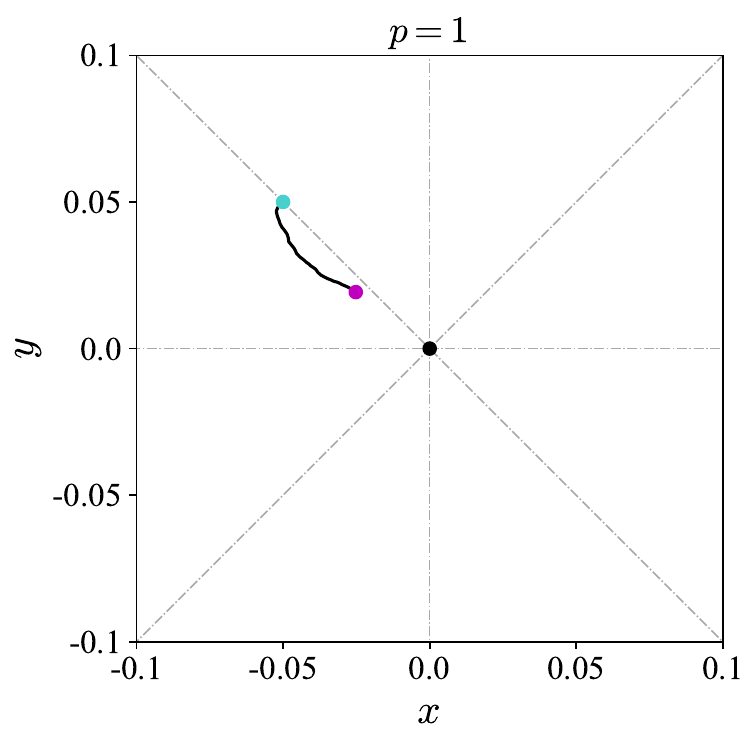}}
	\subfigure{\includegraphics[width=0.4\textwidth]{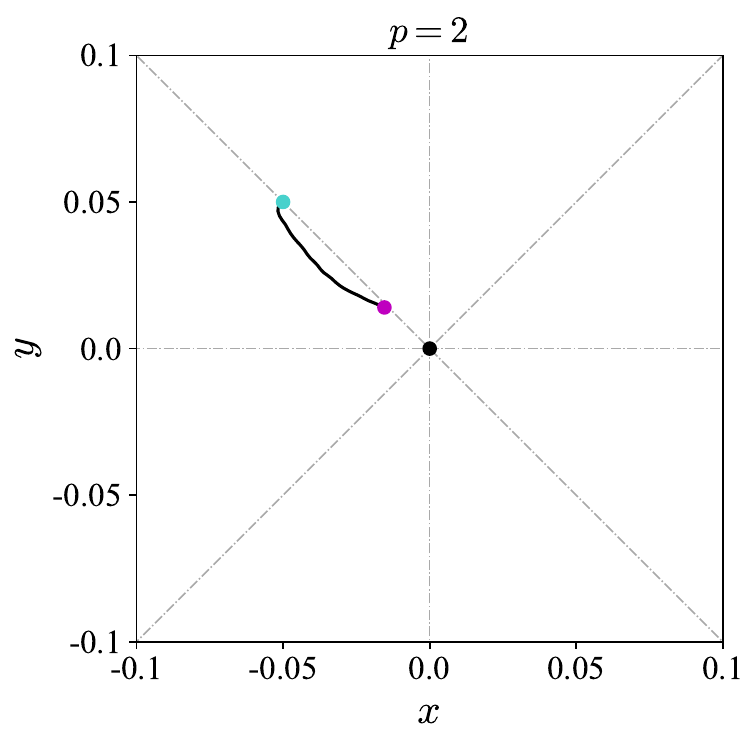}}
	\subfigure{\includegraphics[width=0.4\textwidth]{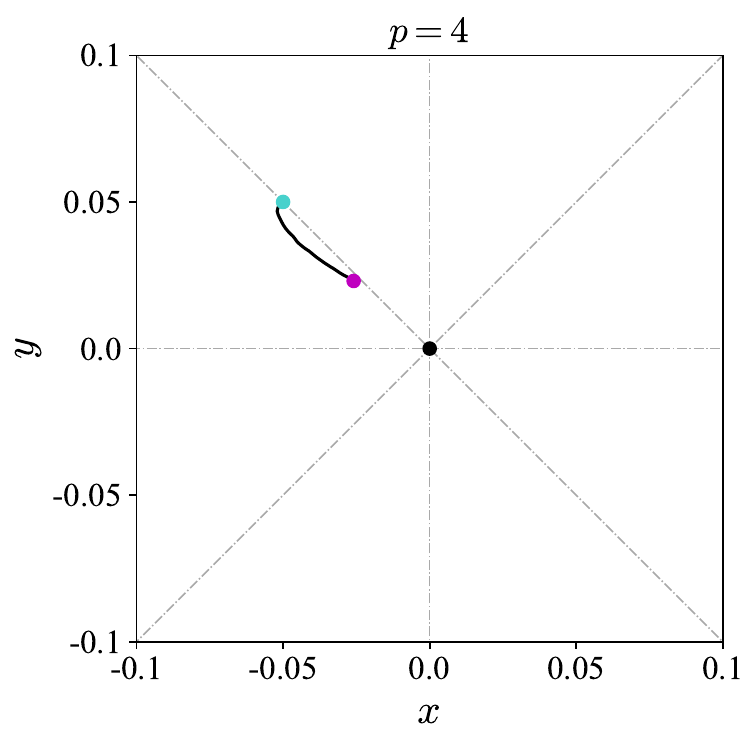}}
	\subfigure{\includegraphics[width=0.4\textwidth]{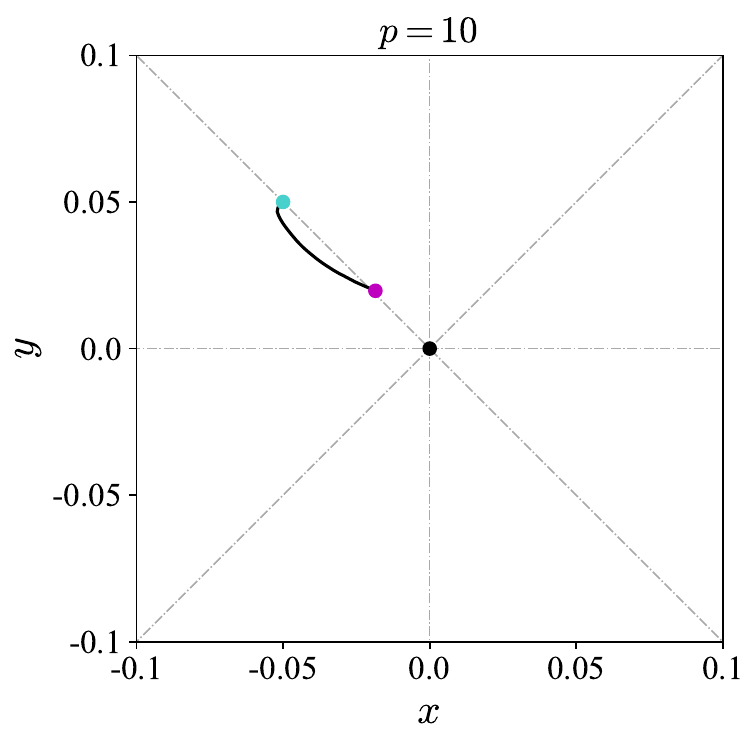}}
	\caption{The trajectories obtained by policies trained using only $r_1(t)$ with different choices of $p$. The other terms in the reward function, i.e., $r_2(t), r'$, are excluded in this test. }
	\label{p_study}
\end{figure}

The overall effect of $p$ in the FRM control is shown in Fig. \ref{p_study}, which displays different trajectories obtained by policies trained by using only $r_1(t)$ with various $p$. The maximum allowable control steps are $N=45$. One can see that $p>1$ results in trajectories closer to the diagonal line which points from the initial point to the origin, consistent with the discussion based on figure \ref{r1_optimal}. For $p\ge2$, no obvious differences are found in terms of the moving direction. In the specific test, the final distance of the $p=2$ case is closest to the target. Therefore, we set $p=2$ in our numerical experiments as the corresponding $r_1(t)$ appears to strike a good balance between exploration and exploitation. 

From these results, we can see that if only $r_1(t)$ is used, the droplet fails to approach closer to the origin. The reason may be that once the droplet moves in the radial direction pointing towards the target, a high reward is given according to $r_1(t)$, which hinders a further improvement of the control. To overcome this, we additionally include $r_2(t)=\exp [-q h(t)]$ which encourages the droplet to move to the target.

\begin{figure}
	\centering
	\subfigure{\includegraphics[width=0.4\textwidth]{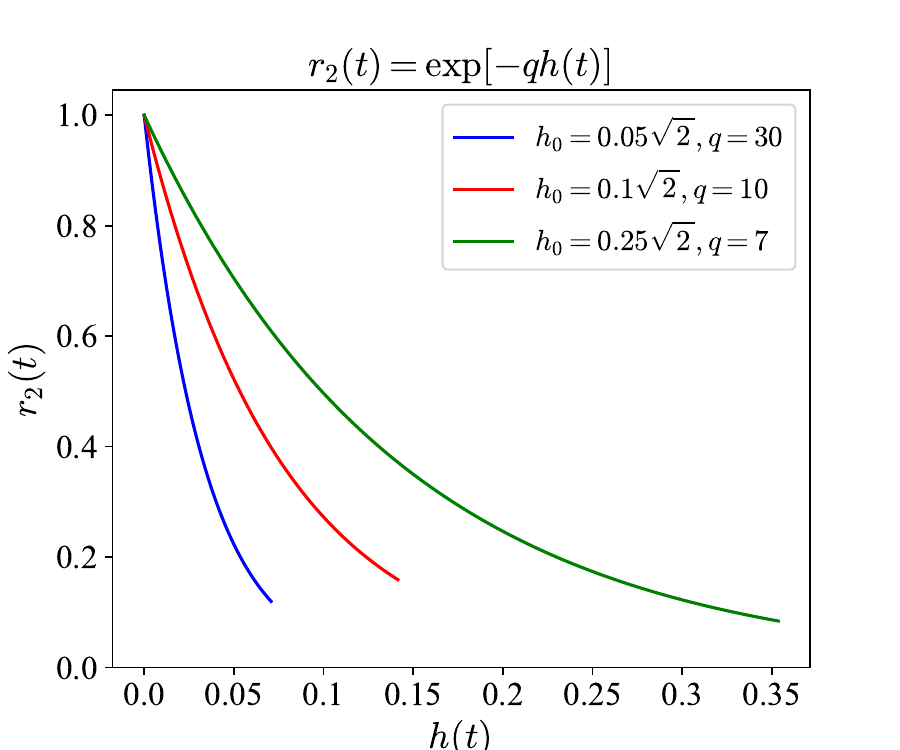}}
	\caption{The second reward term $r_2(t)$ against the distance to the origin $h(t)$.}
	\label{r2_optimal}
\end{figure}

\begin{figure}
	\centering
	\subfigure{\includegraphics[width=0.35\textwidth,trim= 0 0cm 0 0]{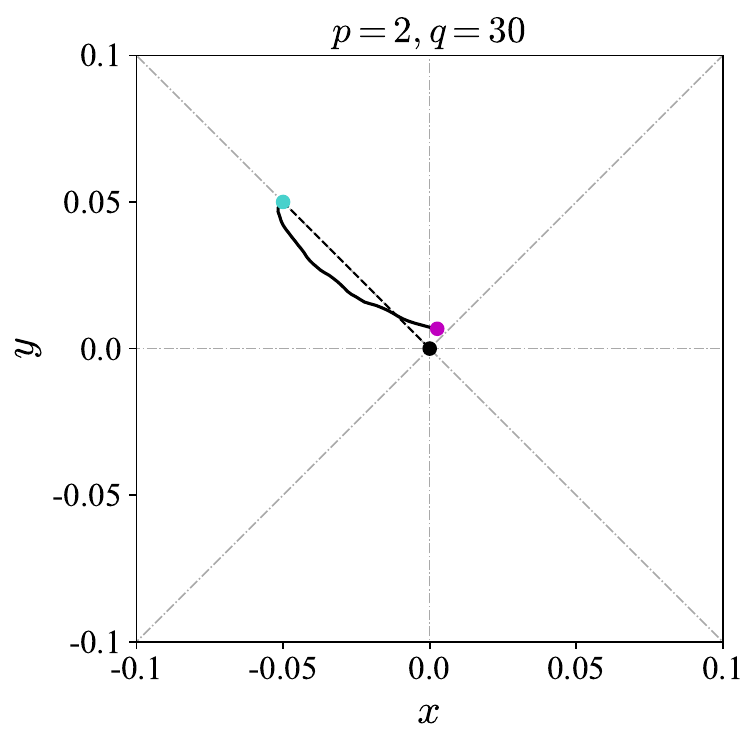}}
	\put(-140,125){$(a)$}
	\hspace{0.2cm}
	\subfigure{\includegraphics[width=0.43\textwidth, trim = 0 0 0 0cm]{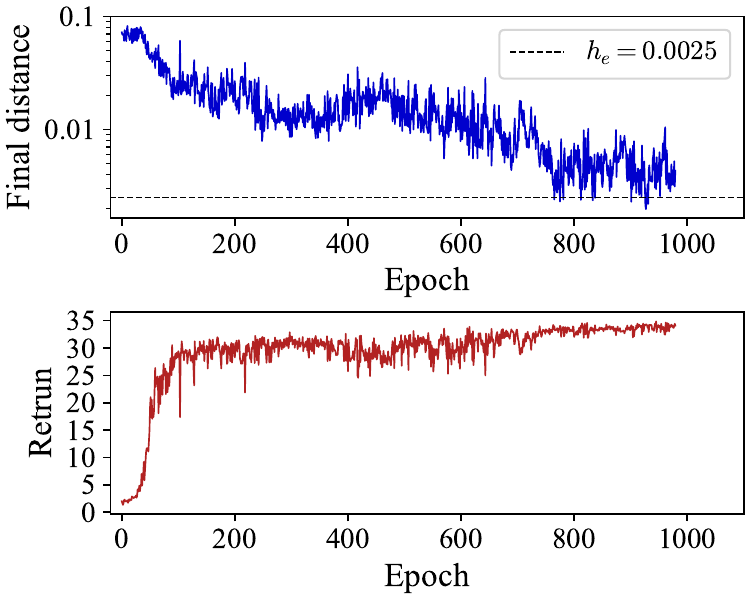}}
	\put(-165,130){$(b)$}\put(-165,67){$(c)$}
	\caption{(a) The trajectory obtained by policy trained with $r_1(t)+r_2(t)'$ using $[p,q]=[2,30]$ without adding $r'$; (bc) The training history.}
	\label{p2_q30}
\end{figure}

\subsection{The ${r_2(t)}$ term} To determine the value of $q$ in $r_2(t)$, we first plot $r_2(t)$ as a function of $h(t)$ for different values of $q$, as shown in figure \ref{r2_optimal}. We aim to design such that the range of $r_2(t)$ corresponds to the values of $h(t)$ which cover the entire distance of the droplet trajectory in the control. Specifically, for $h_0=0.05\sqrt{2}$, the value of $q$ is determined to be $q=30$. 

\mbox{Figure \ref{p2_q30}} displays the trajectory obtained by policy trained using $r_1(t)+r_2(t)$ with $p=2,q=30$, and the associated training history. Compared with those in figure \ref{p_study}, the trajectory in \mbox{figure \ref{p2_q30}} ends up closer to the origin, proving the effectiveness of $r_2(t)$.

\begin{figure}
	\centering
	\subfigure{\includegraphics[width=0.35\textwidth,trim= 0 0cm 0 0]{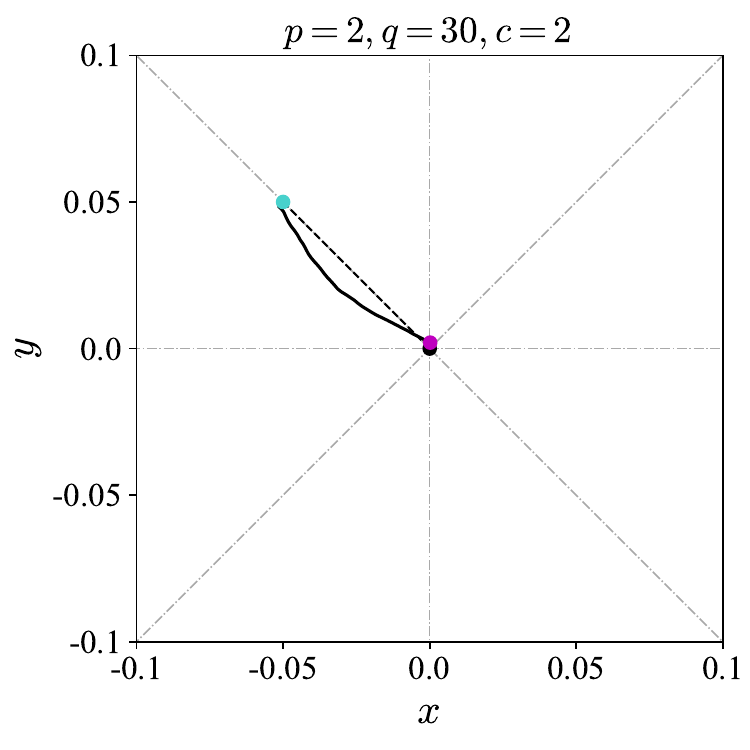}}
	\put(-140,125){$(a)$}
	\hspace{0.2cm}
	\subfigure{\includegraphics[width=0.43\textwidth, trim = 0 0 0 0cm]{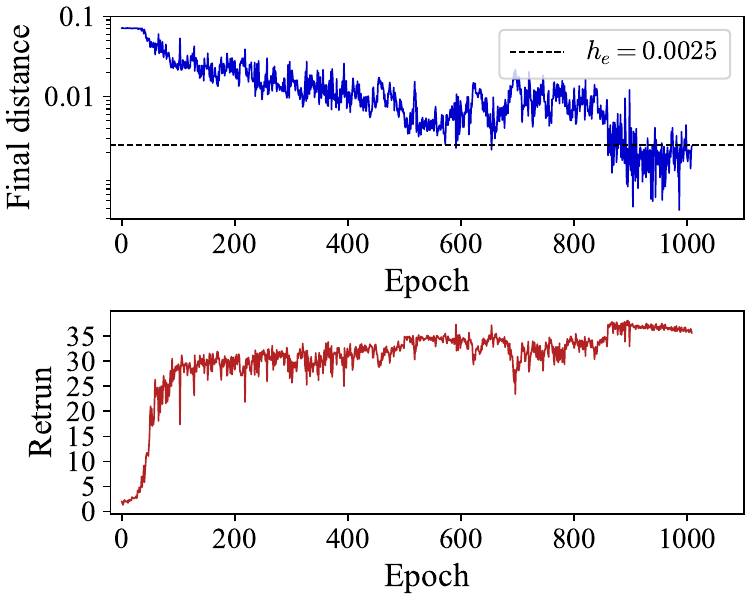}}
	\put(-165,130){$(b)$}\put(-165,67){$(c)$}
	\caption{(a) The trajectory obtained by policy trained with $r_1(t)+r_2(t)+r'$ using $[p,q,c]=[2,30,2]$; (bc) The training history.}
	\label{p2_q30_c2}
\end{figure}

\begin{figure}
	\centering
	\subfigure{\includegraphics[width=0.35\textwidth,trim= 0 0 0 0]{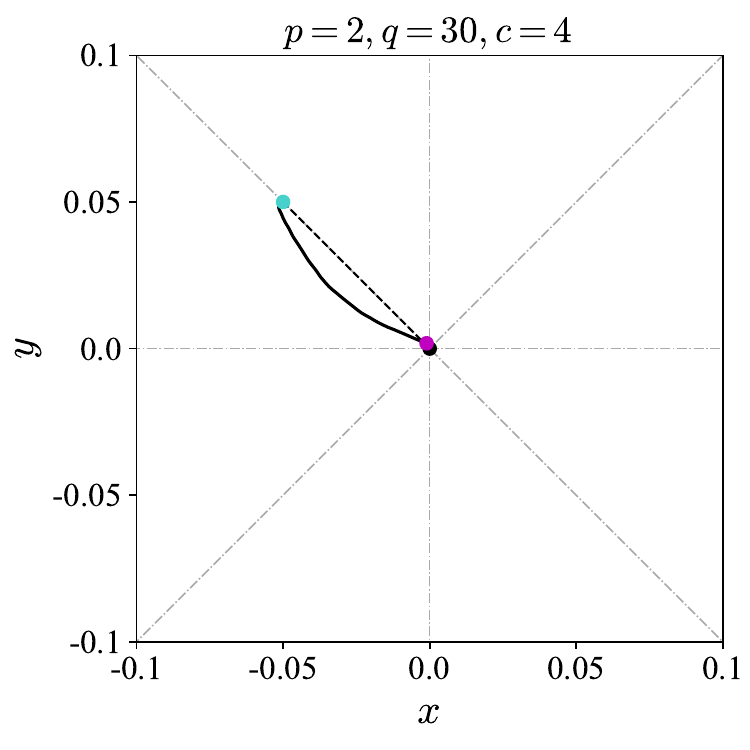}}
	\put(-140,125){$(a)$}
	\hspace{0.2cm}
	\subfigure{\includegraphics[width=0.43\textwidth, trim = 0 0 0 0cm]{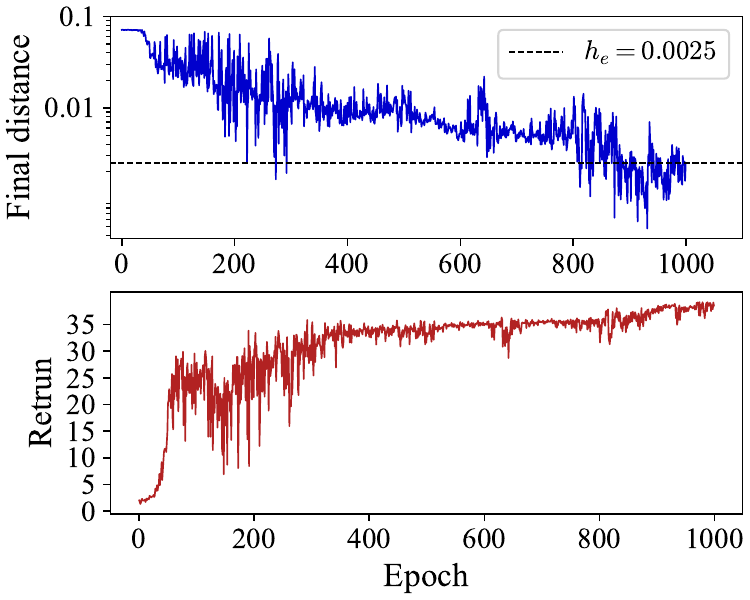}}
	\put(-165,130){$(b)$}\put(-165,67){$(c)$}
	\caption{(a) The trajectory obtained by policy trained with $r_1(t)+r_2(t)+r'$ using $[p,q,c]=[2,30,4]$; (bc) The training history.}
	\label{p2_q30_c4}
\end{figure}

\subsection{The ${r'}$ term} In order to further motivate the agent guiding the droplet to the origin within an expected threshold $h_e$, we add the third reward term $r'= \begin{cases}c, & h(t) \leqslant h_e \\ 0, & h(t)>h_e\end{cases}$. We would like to use a relatively high value for $c$ since its contribution to the return will be repeatedly discounted. And a larger $c$ provides a greater incentive to arrive at the target. In the manuscript, we chose $c=2$ for $h_0=0.05\sqrt{2}$. To verify that this value is effective for the control, we will do a numerical study for $c=2$ and $c=4$. 

\mbox{Figure \ref{p2_q30_c2}} and \mbox{Figure \ref{p2_q30_c4}} display the results with $c=2$ and $c=4$, respecitively. We can see that both trajectories shown in these two cases can reach the target distance within $h_e=0.0025$. Comparing the training history among \mbox{figures \ref{p2_q30}}, \mbox{\ref{p2_q30_c2}} and \mbox{\ref{p2_q30_c4}}, we can see that adding $r'$ improves the convergence to the target distance since this reward term encourages the droplet to move to the origin within prescribed target distance $h_e$. Using $c=4$ only slightly {\color{black} improves} the control performance by termination with fewer epochs. This verifies our choice of $c=2$ in the manuscript. 

\section{\DX{Tests on different state definitions}}\label{statedef}
\begin{figure}
	\centering
	\includegraphics[width=0.33\textwidth]{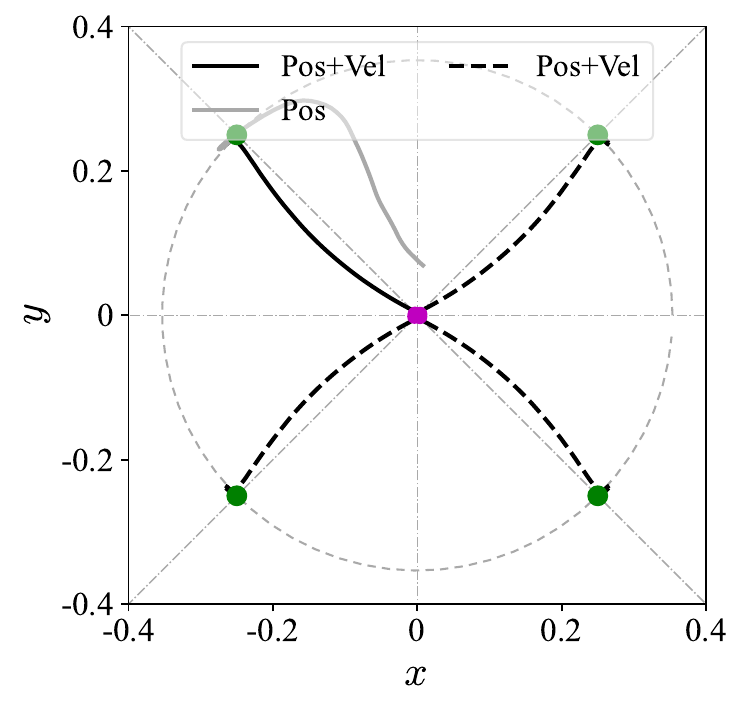}
	\put(-130,115){$(a)$}		
	\includegraphics[width=0.62\textwidth]{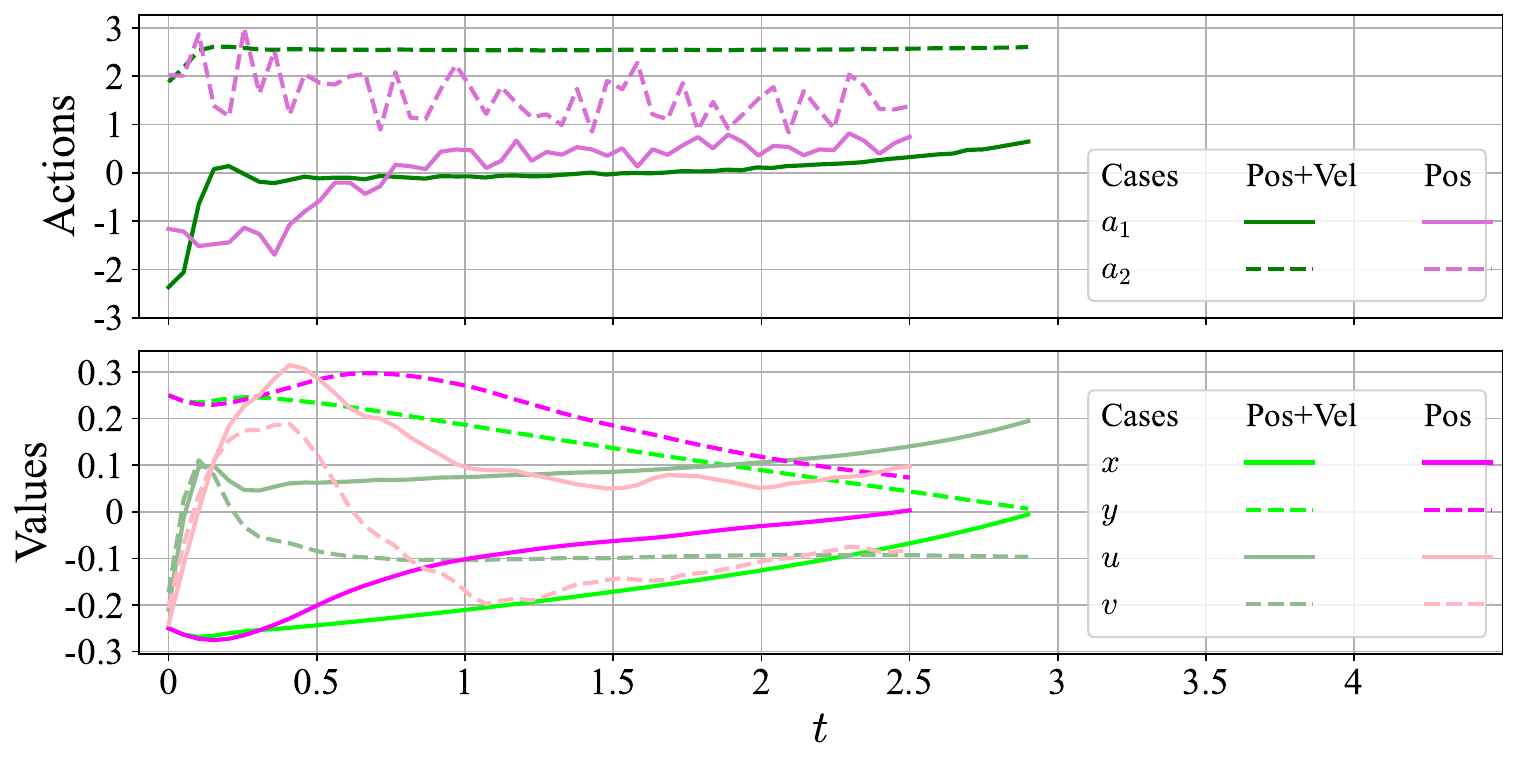}
	\put(-240,115){$(b)$}	
	\caption{($a$) Droplet trajectory under control corresponding to the case {\color{black}5.3} in table \ref{Train_parameters} with the parameters $h_0 = 0.25\sqrt{2}$, $\alpha_0 = 45^{\circ}$ and $Re=2$. The dashed lines result from the application of geometric symmetry in FRM. ($b$) Action history and value history in the test. Additionally, different input configurations for the DRL state were tested, including both position and velocity and position alone. The policy using only position failed to converge. Two closest rollers are activated in the control task.}
	\label{Traj_Re2}
\end{figure}

{In this appendix, {\color{black} we explore a point raised by one of the reviewers that the acceleration in the state definition could be more susceptible to noise compared to other state variables such as position or velocity.} In our numerical method, we simply calculate the acceleration from the time derivative of velocity signals. To test if the acceleration is necessary or not, {\color{black} especially in the case where the acquisition of the acceleration is severely subject to noise}, we provide the following numerical tests.

We focus on the case \(Re = 2\) for illustration. The results for the baseline setup, where the state includes the position, velocity, and acceleration of the droplet, are shown in figure~9(\(a,b\)) in the revised manuscript. In figure~\ref{Traj_Re2} of this document, we successively discard acceleration and velocity from the state definition. By comparing the baseline setup with the case where acceleration is excluded, one can observe that the trajectories are visually identical, although the results for \(u, v\) exhibit slight differences. The actions shown in panel (\(b\)) for these two cases are also highly similar. In both cases, the policies successfully guide the droplet to the target. However, when only position is retained in the state, the control is unsuccessful under otherwise identical parameters. Based on this test, we conclude that the acceleration may not be necessary in our control setup {\color{black}at $Re=2$. Nevertheless, it remains to be tested whether this conclusion holds at higher $Re$, where acceleration may play a more critical role due to stronger inertial effects. However, since significantly increasing $Re$ would disrupt the action-reward relationship in our current DRL algorithm, this investigation will be considered in future work.}
}

\bibliographystyle{jfm}
\bibliography{BibRefs}

\end{document}